\documentclass[useAMS,usenatbib,a4paper]{mn2e}

\usepackage{amssymb}
\usepackage{graphicx}
\usepackage{subfigure}
\usepackage{booktabs}
\usepackage{txfonts}
\usepackage{ifpdf}
\ifpdf
  \usepackage[pdftex]{hyperref}
\else
  \usepackage[ps2pdf,colorlinks=true]{hyperref}
\fi

\newif\ifbw
\bwfalse

\newcommand{\eqn}[1]{(#1)}

\newcommand{\fig}[1]{Fig.~#1}
\newcommand{\Fig}[1]{Fig.~#1}
\newcommand{\sectn}[1]{section~#1}

\newcommand{\etal}{\mbox{\it et al.}}
\newcommand{\eg}{\mbox{\it e.g.}}
\newcommand{\ie}{\mbox{\it i.e.}}

\newcommand{\cmb}{{CMB}}
\newcommand{\cmbtext}{{cosmic microwave background}}

\newcommand{\wmap}{{WMAP}}
\newcommand{\wmaptext}{{Wilkinson Microwave Anisotropy Probe}}

\newcommand{\cobedmr}{\mbox{COBE-DMR}}
\newcommand{\cobedmrtext}{Cosmic Background Explorer-Differential Microwave Radiometer}
\newcommand{\cobe}{\mbox{COBE}}

\newcommand{\isw}{{ISW}}
\newcommand{\iswtext}{{integrated Sachs-Wolfe}}

\newcommand{\lss}{{LSS}}
\newcommand{\lsstext}{large scale structure}

\newcommand{\nvss}{{NVSS}}
\newcommand{\nvsstext}{{NRAO VLA Sky Survey}}

\newcommand{\sdss}{{SDSS}}
\newcommand{\sdsstext}{Sloan Digital Sky Survey}

\newcommand{\twomass}{{2MASS}}
\newcommand{\twomasstext}{Two Micron All Sky Survey}

\newcommand{\healpix}{{\tt HEALPix}}

\newcommand{\lambdaarch}{{LAMBDA}}
\newcommand{\lambdaarchtext}{{Legacy Archive for Microwave Background Data Analysis}}

\newcommand{\kpzero}{{Kp0}}

\newcommand{\spcend}{\ensuremath{\:}}

\newcommand{\cconj}{\ensuremath{\ast}}

\newcommand{\integers}{\ensuremath{\mathbb{Z}}}
\newcommand{\naturals}{\ensuremath{\mathbb{N}}}

\newcommand{\vect}[1]{\ensuremath{\mbox{\boldmath ${#1}$}}}

\newcommand{\opnexpv}{\ensuremath{\langle}}
\newcommand{\clsexpv}{\ensuremath{\rangle}}

\newcommand{\dx}{\ensuremath{\mathrm{\,d}}}

\newcommand{\ctime}{\ensuremath{\eta}}

\newcommand{\z}{\ensuremath{z}}

\newcommand{\sa}{\ensuremath{\omega}}
\newcommand{\saa}{\ensuremath{\theta}}
\newcommand{\sab}{\ensuremath{\varphi}}
\newcommand{\sas}{\ensuremath{\saa, \sab}}

\newcommand{\el}{\ensuremath{\ell}}
\newcommand{\m}{\ensuremath{n}}

\newcommand{\p}{\ensuremath{^\prime}}

\newcommand{\scalea}{\ensuremath{a}}

\newcommand{\kron}[2]{\ensuremath{\delta_{{#1}{#2}}}}

\newcommand{\shf}[2]{\ensuremath{Y_{#1#2}}}

\newcommand{\sbessel}[1]{\ensuremath{j_{#1}}}

\newcommand{\nside}{\ensuremath{{N_{\rm{side}}}}}

\newcommand{\clnttheo}{\ensuremath{C_\el^{\rm NT}}}

\newcommand{\ndlab}{\ensuremath{{\rm N}}}
\newcommand{\tplab}{\ensuremath{{\rm T}}}
\newcommand{\dndz}{\ensuremath{\frac{\dx N}{\dx z}}}
\newcommand{\dgdz}{\ensuremath{\frac{\dx g}{\dx z}}}

\newcommand{\chisqd}{\ensuremath{\chi^2}}

\newcommand{\halfw}{\ensuremath{\theta_{\rm hw}}}
\newcommand{\gpot}{\ensuremath{\Phi}}
\newcommand{\bias}{\ensuremath{b}}

\newcommand{\nd}{\ensuremath{\Delta^{\rm N}}}
\newcommand{\tp}{\ensuremath{\Delta^{\rm T}}}
\newcommand{\tpconj}{\ensuremath{\Delta^{{\rm T} \:\: \cconj}}}

\newcommand{\twogd}{{second derivative of a Gaussian}}
\newcommand{\morphc}{\ensuremath{{D}}}
\newcommand{\morphe}{\ensuremath{{E}}}
\newcommand{\morphi}{\ensuremath{{I}}}
\newcommand{\morphs}{\ensuremath{{S}}}

\title[Probing dark energy with steerable wavelets]
  {Probing dark energy with steerable wavelets through correlation of WMAP and NVSS local morphological measures}

\author[McEwen \etal]
  {J.~D.~McEwen$^{1}$\thanks{E-mail: mcewen@mrao.cam.ac.uk}, Y.~Wiaux$^{2}$, M.~P.~Hobson$^{1}$, P.~Vandergheynst$^{2}$, A.~N.~Lasenby$^{1}$\\ 
  $^1$Astrophysics Group, 
      Cavendish Laboratory,  J.~J.~Thomson Avenue,
      Cambridge CB3 0HE, UK\\
  $^2$Signal Processing Institute, Ecole Polytechnique F\'ed\'erale de Lausanne (EPFL), 
      CH-1015 Lausanne, Switzerland\\
}

\date{Accepted 27 November 2007. Received 27 November 2007; in original form 12 April 2007}
\pagerange{\pageref{firstpage}--\pageref{lastpage}} 
\pubyear{2007}

\def\LaTeX{L\kern-.36em\raise.3ex\hbox{a}\kern-.15em
    T\kern-.1667em\lower.7ex\hbox{E}\kern-.125emX}

\begin{document}
\label{firstpage}
\maketitle


\begin{abstract}
Using local morphological measures on the sphere defined through a steerable wavelet analysis, we examine the three-year \wmap\ and the \nvss\ data for correlation induced by the \iswtext\ (\isw) effect.  
The steerable wavelet constructed from the second derivative of a Gaussian allows one to define three local morphological measures, namely the signed-intensity, orientation and elongation of local features.  Detections of correlation between the \wmap\ and \nvss\ data are made with each of these morphological measures.  The most significant detection is obtained in the correlation of the signed-intensity of local features at a significance of 99.9\%.  
By inspecting signed-intensity sky maps, it is possible for the first time to see the correlation between the \wmap\ and \nvss\ data by eye.
Foreground contamination and instrumental systematics in the \wmap\ data are ruled out as the source of all significant detections of correlation.  
Our results provide new insight on the \isw\ effect by probing the morphological nature of the correlation induced between the \cmbtext\ and \lsstext\ of the Universe.  Given the current constraints on the flatness of the Universe, our detection of the \isw\ effect again provides direct and independent evidence for dark energy.  Moreover, this new morphological analysis may be used in future to help us to better understand the nature of dark energy.
\end{abstract}


\begin{keywords}
cosmic microwave background -- cosmology: observations -- methods: data analysis -- methods: numerical.
\end{keywords}

\section{Introduction}

A cosmological concordance model has emerged recently, explaining many observations of our Universe to very good approximation.  In this model our Universe is dominated by an exotic dark energy component that may be described by a cosmological fluid with negative pressure, interacting only gravitationally to counteract the attractive gravitational nature of matter.  Dark energy may be strongly inferred from observations of the \cmbtext\ (\cmb), such as the \wmaptext\ (\wmap) data \citep{bennett:2003a,hinshaw:2006}, together with either observations of Type Ia Supernovae \citep{riess:1998,perlmutter:1999} or of the \lsstext\ (\lss) of the Universe (\eg\ \citealt{allen:2002}).  
Although dark energy dominates the energy density of our Universe, we know very little about its origin and nature.  Indeed, a consistent model to describe dark energy in the framework of particle physics is lacking.

The \iswtext\ (\isw) effect \citep{sachs:1967} provides an independent physical phenomenon that may be used to detect and probe dark energy.  As \cmb\ photons travel towards us from the surface of last scattering they pass through gravitational potential wells due to the \lss.  If the gravitational potential evolves during the photon propagation, then the photon undergoes an energy shift.  The \isw\ effect is the integrated sum of this energy shift along the photon path.  In a matter-dominated Einstein-de Sitter universe the gravitational potential remains constant with respect to conformal time, thus there is no \isw\ effect.  However, if the universe deviates from matter domination due to curvature or dark energy then an \isw\ effect is induced.  Strong constraints have been placed on the flatness of the Universe by \wmap\ \citep{spergel:2006}, hence a detection of the \isw\ effect may be inferred as direct and independent confirmation of dark energy.

It is difficult to isolate the \isw\ contribution to \cmb\ anisotropies directly, therefore \citet{crittenden:1996} suggested that the \isw\ effect be detected by cross-correlating \cmb\ anisotropies with tracers of the local matter distribution, such as the nearby galaxy density distribution.  A positive large-scale correlation will be induced by the \isw\ effect as a consequence of decaying gravitational potentials due to dark energy.  First attempts to detect the \isw\ effect using \cobedmrtext\ (\cobedmr) data failed, concluding that greater sensitivity and resolution than that provided by \cobe\ were required \citep{boughn:2002}.  Fortunately, the \wmap\ mission soon provided suitable \cmb\ data.  Correlations indicative of the \isw\ effect have now been detected between both the first- and three-year \wmap\ data and a large number of tracers of the \lss.  
Detections have been made using
\nvsstext\ (\nvss; \citealt{condon:1998}) data \citep{boughn:2004,nolta:2004,vielva:2005,mcewen:2006:isw,pietrobon:2006}, 
hard X-ray data provided by the High Energy Astronomy Observatory-1 satellite (HEAO-1; \citealt{boldt:1987}) \citep{boughn:2005,boughn:2004},
APM galaxy survey \citep{maddox:1990} data \citep{fosalba:2004},
\sdsstext\ (\sdss; \citealt{york:2000}) data  \citep{scranton:2003,fosalba:2003,padmanabhan:2004,cabre:2006,giannantonio:2006}, 
the \twomasstext\ Extended Source Catalogue (\twomass\ {XSC}; \citealt{jarrett:2000}) \citep{afshordietal:2004,rassat:2006}
and combinations of the aforementioned data sets \citep{gaztanaga:2006}.
Furthermore, a number of other works have focused on theoretical detectability, future experiments and/or error analyses related to \isw\ detections \citep{afshordi:2004,hu:2004,pogosian:2004,pogosian:2006,pogosian:2005,corasaniti:2005,loverde:2006,cabre:2007}.

Many different analysis techniques have been employed in these previous works to detect correlations between the data sets, each of which have their own merits and limitations.  In this work we focus on wavelet-based techniques.  Wavelets provide an ideal analysis tool to search for the \isw\ effect.  This is due to the localised nature of the effect, in both scale and position on the sky, and the simultaneous scale and position localisation afforded by a wavelet analysis.  Since the \isw\ effect is cosmic variance limited, it is desirable to examine as great a sky coverage as possible.  In this near full-sky setting the geometry of the sphere should be taken into account, thus wavelet analyses on the sphere are required.  
The first analysis using wavelets on the sphere to search for correlations between the \cmb\ and \lss\ was performed by \citet{vielva:2005} using the axisymmetric spherical Mexican hat wavelet.  \citet{mcewen:2006:isw} extended this analysis to directional wavelets, as correlated features induced by the \isw\ effect may not necessarily be rotationally invariant.  Indeed, it is known that statistically isotropic Gaussian random fields are characterised by local features that are not rotationally invariant \citep{barreiro:1997,barreiro:2001}.  Recently, \citet{pietrobon:2006} applied a new wavelet construction on the sphere called needlets to search for correlations.  All of these works examined the \wmap\ and \nvss\ data (due to the large sky coverage of the \nvss\ data) and made significant detections of the \isw\ effect.  Moreover, in each analysis the detection of the \isw\ effect was used to constrain dark energy parameters.

In this paper we present a new search for correlation between the three-year \wmap\ and the \nvss\ data using wavelets on the sphere.  However, the approach taken here differs fundamentally to previous wavelet analyses.  We use steerable wavelets on the sphere to extract, from each of the two data sets, measures of the morphology of local features, such as their signed-intensity, their orientation, or their elongation.  We then correlate the \wmap\ and \nvss\ data through these quantities.  This represents the first direct analysis of local morphological measures on the sphere resulting from the steerability of wavelets.  A positive detection of correlation using any of these morphological probes would give a direct indication of the \isw\ effect, provided it were not due to unremoved foregrounds or measurement systematics, which could again be inferred as direct and independent evidence for dark energy.
Moreover, such a detection would give an indication of the morphological nature of the correlation.  This further insight might help to better understand the nature of dark energy.  However, the derivation of the theoretical correlation of the newly defined morphological measures as a function of the dark energy content of the Universe is not easily tractable.  Consequently, corresponding constraints on dark energy parameters are left to a future work.

The remainder of this paper is organised as follows.  In \sectn{\ref{sec:background}} we present the physical and signal processing background behind the analyses performed.  The \isw\ effect is described and the use of local morphological measures defined through a steerable wavelet analysis is presented as a probe to search for the \isw\ effect.  In \sectn{\ref{sec:procedures}} the data and analysis procedures employed are described in detail.  Two distinct analysis techniques are proposed to compute correlations from morphological measures: a \emph{local morphological analysis} and a \emph{matched intensity analysis}.  The results of these analyses are presented in \sectn{\ref{sec:morph}} and \sectn{\ref{sec:intensity}} respectively.  Concluding remarks are made in \sectn{\ref{sec:conclusions}}.

\section{\isw\ effect and steerable wavelets}
\label{sec:background}

The existence of an \isw\ effect induces a cross-correlation
between the \cmb\ and \nvss\ signals on the celestial sphere. From the
spectral point of view, the corresponding two-point cross-correlation
function may be expressed in terms of the cross-correlation angular
power spectrum. In this paper, we go beyond this spectral
analysis thanks to a decomposition of the \cmb\ and \nvss\ signals with
steerable wavelets on the sphere. The wavelet analysis enables one to
probe local features at each analysis scale. 
%
In this section we first review the \isw\ effect and the correlation between the \cmb\ and \nvss\ data that it induces.  Secondly, we discuss wavelets on the sphere and describe how the steerability of wavelets is used to define local morphological measures on the sphere.  

\subsection{\isw\ effect}
\label{sec:isw}

The secondary temperature anisotropy induced in the \cmb\ by the \isw\ effect is related to the evolution of the gravitational potential.  Any recent acceleration of the scale factor of the Universe due to dark energy will cause local gravitational potentials to decay.  \cmb\ photons passing through over dense regions of decaying potential suffer blue shifts, resulting in a positive correlation between the induced anisotropy and the local matter distribution.  It can be shown that the relative temperature fluctuation induced in the \cmb\ is given by
\begin{equation}
\label{eqn:isw}
\tp(\sa) =  \frac{\Delta T(\sa)}{T_0} = -2 \int \dx \ctime \: \dot{\gpot}(\ctime,\sa)
\end{equation}
\citep{sachs:1967,nolta:2004},
where $\Delta T$ is the induced temperature perturbation, $T_0$ is the mean temperature of the \cmb, \ctime\ is conformal time, \gpot\ is the gravitational potential and the dot represents a derivative with respect to conformal time.  
A point \sa\ on the sky is represented in spherical coordinates as $\sa=(\sas)$, with co-latitude \saa\ and longitude \sab.
The integral is computed over the photon path from emission to observation, \ie\ from today back to the last scattering surface.
In a matter-dominated Einstein-de Sitter universe (with zero cosmological constant) the potential evolves as $\gpot \sim \delta / R$, where $\delta$ is the matter perturbation and $R$ is the scale factor of the universe.  In this setting the matter perturbation evolves with the scale factor, $\delta \propto R$. Consequently, $\dot{\gpot}=0$ and there is no \isw\ effect, as discussed previously.

We use the \nvss\ galaxy count distribution projected onto the sky as a tracer of the local matter distribution.  It is assumed that the two corresponding relative fluctuations, respectively $\delta^\ndlab(\z,\sa)$ and $\delta(\z,\sa)$, are related by the linear bias factor $\bias(\z)$: \mbox{$\delta^\ndlab(\z,\sa) = \bias(\z) \, \delta(\z,\sa)$}, where \z\ is redshift.  Hereafter we take the bias to be redshift independent since the redshift epoch 
over which the \isw\ effect is produced is small.
The galaxy source count fluctuation observed on the sky is therefore given by
\begin{equation}
\label{eqn:galaxy}
\nd (\sa) = \bias \: \int \dx z \: \frac{\dx N}{\dx \z} \: \delta(\z, \sa)
\spcend ,
\end{equation}
where $\dx N / \dx \z$ is the mean number of sources per steradian at redshift \z\ and the integral is performed from today to the epoch of recombination, \ie\ last scattering.

We are now in a position to consider the correlation between the galaxy count and \cmb\ temperature fluctuations.  We consider the cross-power spectrum \clnttheo\ defined by the ensemble average of the product of the spherical harmonic coefficients of the two signals observed on the sky:
\begin{equation}
\label{eqn:cltheo}
\opnexpv \nd_{\el\m} \: \tpconj_{\el\p \m\p} \clsexpv =
\kron{\el}{\el\p} \kron{\m}{\m\p} \:
\clnttheo
\spcend ,
\end{equation}
where 
$\Delta_{\el\m} = \opnexpv \shf{\el}{\m} | \Delta \clsexpv$ are the spherical harmonic coefficients of $\Delta(\sa)$, $\opnexpv \cdot | \cdot \clsexpv$ denotes the inner product on the sphere, $\shf{\el}{\m}$ are the spherical harmonic functions for multipole $\el\in\naturals$, $\m\in\integers$, $|\m|\leq\el$ and $\kron{i}{j}$ is the Kronecker delta symbol.  In writing the cross-correlation in this manner we implicitly assume that the galaxy density and \cmb\ random fields on the sphere are homogeneous and isotropic, which holds under the basic assumption of the cosmological principle.  Representing the gravitational potential and the matter density perturbation in Fourier space and substituting \eqn{\ref{eqn:isw}} and \eqn{\ref{eqn:galaxy}} into \eqn{\ref{eqn:cltheo}}, it is possible to show (\eg\ \citealt{nolta:2004}) that 
\begin{equation}
\clnttheo = 12 \pi \: \Omega_{\rm m} \: H_0{}^2
\int \frac{\dx k}{k^3} \:
\Delta_\delta^2(k) \:
F_\el^{\rm N}(k) \:
F_\el^{\rm T}(k)
\spcend ,
\end{equation}
where $\Omega_{\rm m}$ is the matter density, $H_0$ is the Hubble parameter, \mbox{$\Delta_\delta^2(k) = k^3 P_\delta(k)/2\pi^2$} is the logarithmic matter power spectrum,
$P_\delta(k)=\opnexpv |\delta(k)|^2 \clsexpv$ is the matter power spectrum and
the filter functions for the galaxy density and \cmb\ are given by
\begin{equation}
F_\el^{\ndlab}(k) = b \int \dx z \: \dndz \: D(z) \: \sbessel{\el}[k \eta(z)]
\end{equation}
and
\begin{equation}
F_\el^{\tplab}(k) = \int \dx z \: \dgdz \: \sbessel{\el}[k \eta(z)]
\end{equation}
respectively.  The integration required to compute $F_\el^\tplab(k)$ is performed over $z$ from zero to the epoch of recombination, whereas, in practice, the integration range for $F_\el^\ndlab(k)$ is defined by the source redshift distribution $\dx N / \dx \z$.  $D(\z)$ is the linear growth factor for the matter distribution: $\delta(\z,k)=D(\z) \delta(k)$, with $\delta(k) = \delta(0,k)$.  The function $g(\z) \equiv (1+\z)D(\z)$ is the linear growth suppression factor and $\sbessel{\el}(\cdot)$ is the spherical Bessel function.
We have represented in harmonic space the expected correlation between the galaxy source count fluctuations and the \cmb\ temperature fluctuations induced by the \isw\ effect.  We next turn our attention to steerable wavelets on the sphere as a potential tool for detecting this correlation.

\subsection{Wavelets on the sphere}

The \wmap\ and \nvss\ data may be understood as the sampling of continuous
signals on the sphere, which can be analysed in the framework of a
continuous wavelet formalism.  
The analysis of a signal on the sphere with a wavelet, which is a local analysis function, yields a set of wavelet coefficients. These coefficients result from the scalar products between the signal and the wavelet
dilated at any scale, rotated around itself by any angle, and translated at any point on the sphere. The so-called steerable wavelets allow,
from these wavelet coefficients, the definition of local morphological measures of the signal. In particular, the second Gaussian
derivative wavelet used in the present analysis gives access to  the measures of orientation, signed-intensity, and elongation
of the signal's local features. The remainder of this section is devoted to the explicit definition of these local morphological measures. 
The reader not directly interested in these formal and technical details may proceed directly to \sectn{\ref{sec:procedures}}.

The continuous wavelet formalism on
the sphere originally proposed by \citet{antoine:1998} was recently
further developed in a practical approach by \citet{wiaux:2005}.
It defines the wavelet decomposition of a signal on the sphere in
the following way. We consider an orthonormal Cartesian coordinate
system $(o,o\hat{x},o\hat{y},o\hat{z})$ centred on the (unit) sphere,
with the direction $o\hat{z}$ defining the North Pole. 
To relate this coordinate system to the spherical coordinates $\sa=(\sas)$ defined previously, we let 
the polar angle, or co-latitude, $\theta\in[0,\pi]$ represent the
angle between the vector identifying $\omega$ and the axis $o\hat{z}$,
and the azimuthal angle, or longitude, $\varphi\in[0,2\pi)$ represent
the angle between the orthogonal projection of this vector in the
plane $(o,o\hat{x},o\hat{y})$ and the axis $o\hat{x}$.

The signal to be analysed is represented by a square-integrable function
$F(\omega)$ on the sphere, \ie\ $F(\omega) \in L^{2}(S^{2},\dx\Omega)$,
for the invariant measure $\dx\Omega=\dx(\cos\theta) \dx\varphi$. 
A square-integrable analysis function $\Psi(\omega)$ is defined,
the so-called mother wavelet, initially centred at the North Pole.
This wavelet may be dilated at any scale $a>0$. 
The dilation of a wavelet on the sphere in this formalism may be uniquely related to the usual dilation in the tangent plane at the north pole (as discussed in more detail below).
The corresponding
$\Psi_{a}$ can cover arbitrarily small or large regions of the sphere, respectively corresponding to high or low frequencies.
Through the appropriate three-dimensional rotation, the wavelet may also be rotated on itself by any angle $\chi\in[0,2\pi)$
and translated to any point $\omega_{0}=(\theta_{0},\varphi_{0})$
on the sphere.  The final dilated, rotated and translated wavelet is denoted $\Psi_{\omega_{0},\chi,a}$. 
At each
scale, the wavelet coefficients of the signal are defined by the directional
correlation with $\Psi_{a}$, \ie\ by the simple scalar product 
\begin{equation}
W_{\Psi}^{F}\left(\omega_{0},\chi,a\right)=\langle\Psi_{\omega_{0},\chi,a}\vert F\rangle\equiv\int_{S^{2}}\dx\Omega\,\Psi_{\omega_{0},\chi,a}^{*}\left(\omega\right)F\left(\omega\right)\spcend,
\label{eqn:waveletcoefficients}
\end{equation}
where the superscript $^{*}$ denotes complex conjugation. These
wavelet coefficients hence characterise the signal locally around
each point $\omega_{0}$, at scale $a$ and orientation $\chi$. 
The choice of possible mother wavelets is submitted to a condition
ensuring that the original signal can be reconstructed exactly
from its wavelet coefficients. In practice, this wavelet admissibility condition
on the sphere is not easy to check for a given
candidate function. It is therefore difficult to build wavelets directly
on the sphere.

\begin{figure}
\begin{centering}\includegraphics[width=8cm]{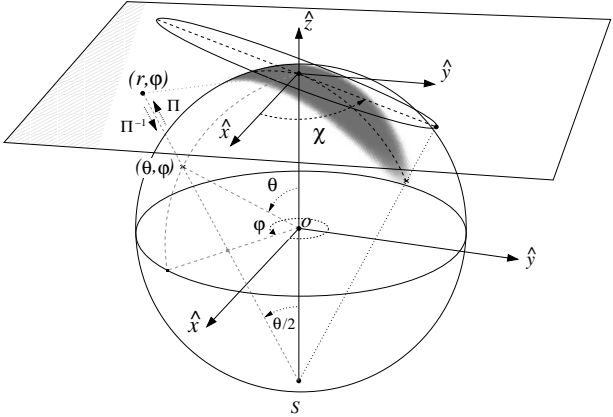}\par\end{centering}
\caption{\label{cap:Stereographic-projection}Wavelets on the sphere can be
defined through simple inverse stereographic projection $\Pi^{-1}$
of wavelets on the tangent plane at the North pole. The corresponding
functions are illustrated here by the shadow on the sphere and the
localised region on the plane.  (Illustration reproduced from \citealt{wiaux:2005}.)
}
\end{figure}

A correspondence principle has been established \citep{wiaux:2005}, which states that the
inverse stereographic projection of a wavelet on the plane provides
a wavelet on the sphere. The stereographic projection of a point $\omega=(\theta,\varphi)$
on the sphere gives by definition a point $\vec{x}=(r,\varphi)$ on
its tangent plane at the North pole, which is co-linear with $\omega$
and the South Pole. It relates the radial variables on the two manifolds
through the bijection $r(\theta)=2\tan(\theta/2)$, and simply identifies
the angular variables $\varphi$ (see \fig{\ref{cap:Stereographic-projection}}).
If the corresponding unitary operator on functions is denoted $\Pi$,
the correspondence principle states that, if $\psi(\vec{x})$ is a
wavelet on the plane, then $\Psi(\omega)=[\Pi^{-1}\psi](\omega)$
is a wavelet on the sphere.  

The dilation of a wavelet on the sphere actually corresponds to the conjugate, through the stereographic projection, of the usual dilation in the tangent plane at the north pole.
This conjugation relation also trivially holds for the rotation of wavelets by $\chi$, which   
only affects the longitude, while the stereographic projection leaves it invariant.
In other words, the dilation and rotation of the wavelet on itself may be performed through the natural dilation and rotation
on the plane before projection:
\begin{equation}
\Psi_{\chi,a}(\omega)=[\Pi^{-1}\psi_{\chi,a}](\omega)
\spcend .
\end{equation}
Only the translation by $\omega_{0}$ must be performed explicitly
on the sphere through the appropriate rotation in three dimensions.
As wavelets on the plane are well-known and may be easily built as
simple zero-mean filters, the correspondence principle enables one to
define a large variety of wavelets on the sphere by simple projection.
This principle also allows one to transfer wavelet properties from the
plane onto the sphere, such as the notion of steerability discussed
in the next section.

The wavelet analysis on the sphere described above is computationally demanding.  The directional correlation described by \eqn{\ref{eqn:waveletcoefficients}} requires the evaluation of a two-dimensional integral over a three-dimensional domain.  Direct computation of this integral by simple quadrature is not computationally feasible for practical \cmb\ data, such as the currently available 3 megapixel \wmap\ maps and the forthcoming 50 megapixel Planck \citep{planck:bluebook} maps.  Fast algorithms to compute the wavelet decomposition of signals on the sphere have been developed and implemented by \citet{wiaux:2005c} and \citet{mcewen:2006:fcswt} (for a review see \citealt{wiaux:2006:review}).  In this work we utilise the implementation described by \citet{mcewen:2006:fcswt}.

\subsection{Steerability and morphological measures}
\label{sec:steerability}

Firstly, we review the notion of steerable wavelets on the sphere. Just
as on the plane, a function on the sphere is said to be axisymmetric
if it is invariant under rotations around itself. Any non-axisymmetric
function is called directional. The directionality of a wavelet is
essential in allowing one to probe the direction of local features in
a signal. A wavelet is said to be steerable if its rotation around
itself by any angle $\chi$ may be written as a simple linear combination
of non-rotated basis filters. On the sphere, for $M$ 
basis filters $ \Psi_{m}$, a steerable wavelet may be written, by definition, as
\begin{equation}
\Psi_{\chi}(\omega)=\sum_{m=1}^{M}k_{m}(\chi)\Psi_{m}(\omega)
\end{equation}
\citep{wiaux:2005}.
The weights $k_{m}(\chi)$, with $1\leq m\leq M$, are called interpolation
functions. In particular cases, the basis filters may be specific
rotations by angles $\chi_{m}$ of the original wavelet: $\Psi_{m}=\Psi_{\chi_{m}}$.
Steerable wavelets have a non-zero angular width in the azimuthal angle
$\varphi$. This property makes them sensitive to a range of
directions and enables them to satisfy their relation of definition.
In spherical harmonic space, this non-zero angular width corresponds
to an azimuthal angular band limit $N$ in the integer index $n$
associated with the azimuthal variable $\varphi$: ${\Psi}_{\el\m}=0$
for $\vert n\vert\geq N$.
Typically, the number $M$ of interpolating functions is of the same
order as the azimuthal band limit $N$. Notice that the steerability
is a property related to the azimuthal variable $\varphi$ only. Again as
the stereographic projection only affects the radial variables, steerable
wavelets on the sphere can easily be built as simple inverse stereographic
projection of steerable wavelets on the plane.

\begin{figure}
\centering
\subfigure[$\Psi^{\partial_{\hat{x}}^{2}(Gauss)}$]{\includegraphics[width=2.5cm]{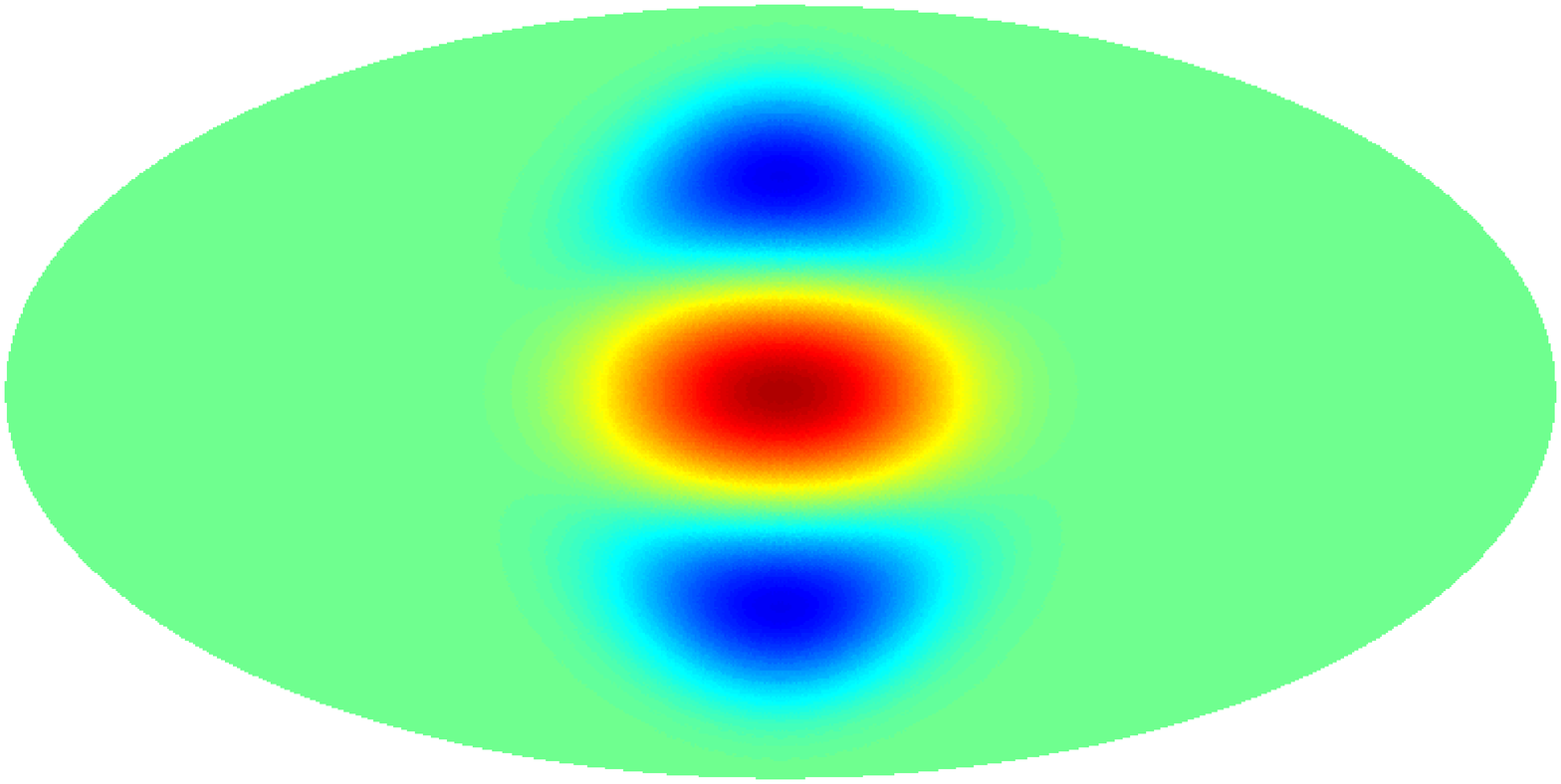}} \quad
\subfigure[$\Psi^{\partial_{\hat{y}}^{2}(Gauss)}$]{\includegraphics[width=2.5cm]{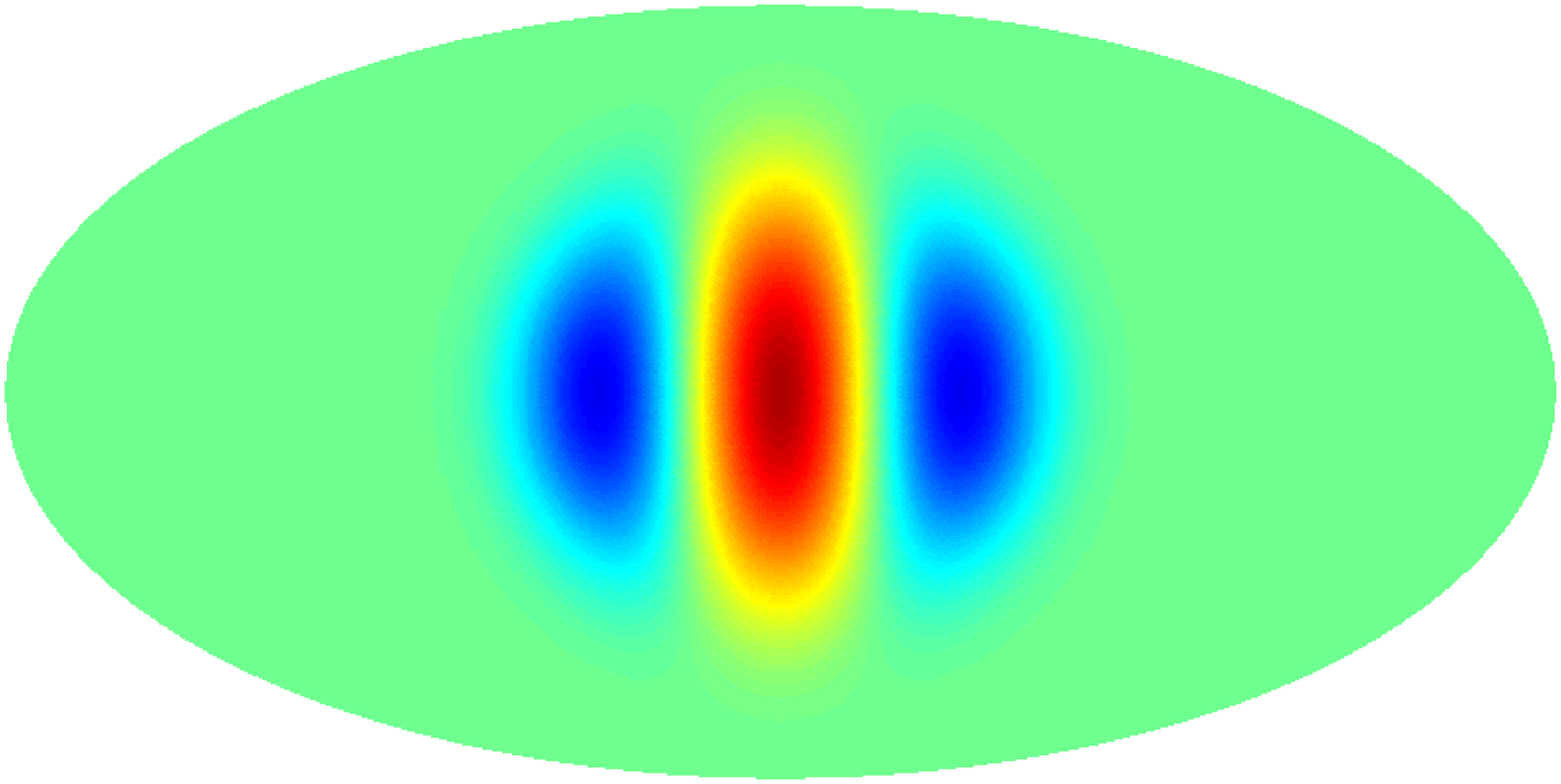}} \quad
\subfigure[$\Psi^{\partial_{\hat{x}}\partial_{\hat{y}}(Gauss)}$]{\includegraphics[width=2.5cm]{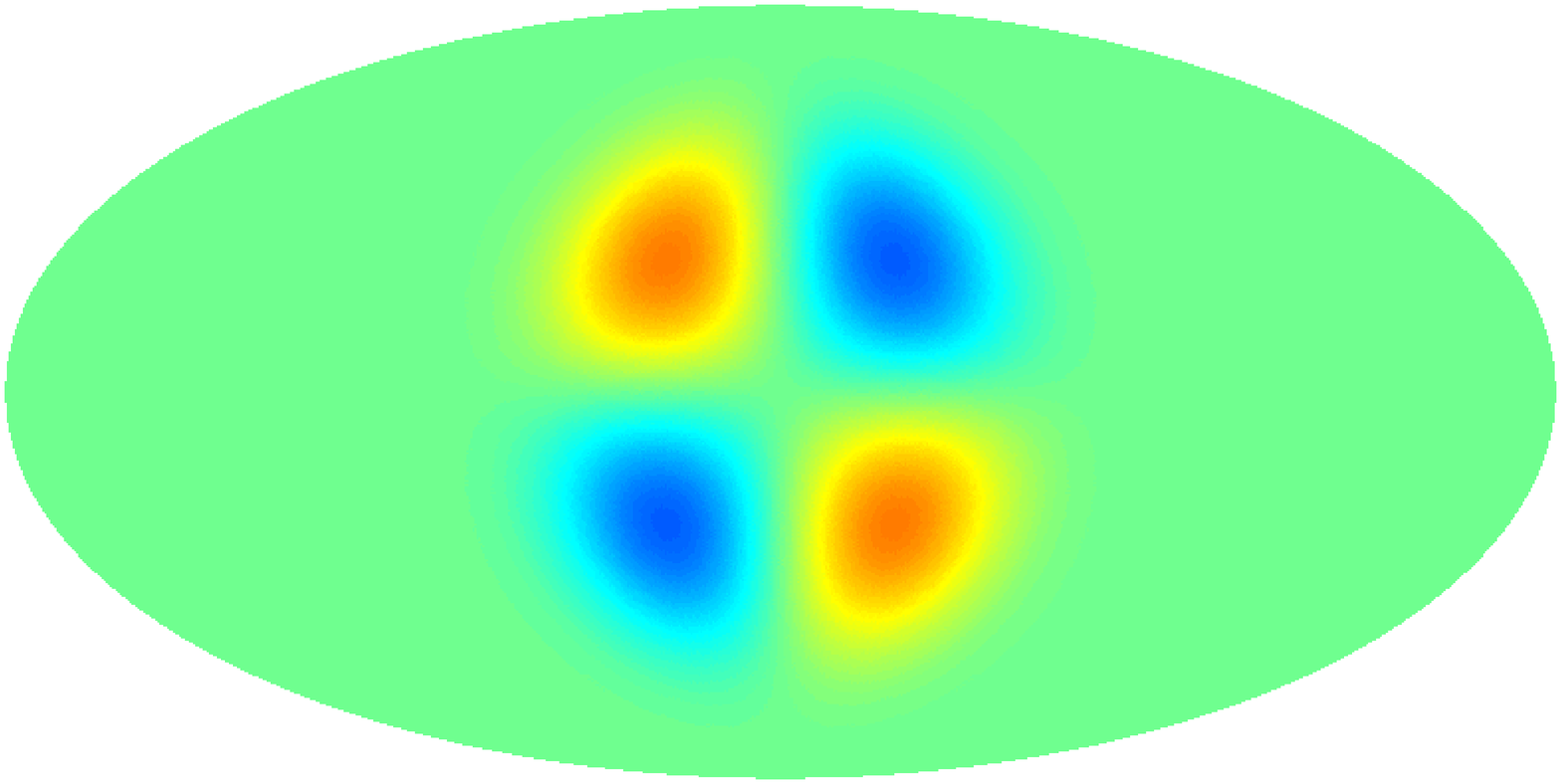}} \\
\subfigure[$\Psi^{\partial_{\hat{x}}^{2}(Gauss)}$ rotated by $\chi=\pi/4$]{\hspace*{10mm}\includegraphics[width=2.5cm]{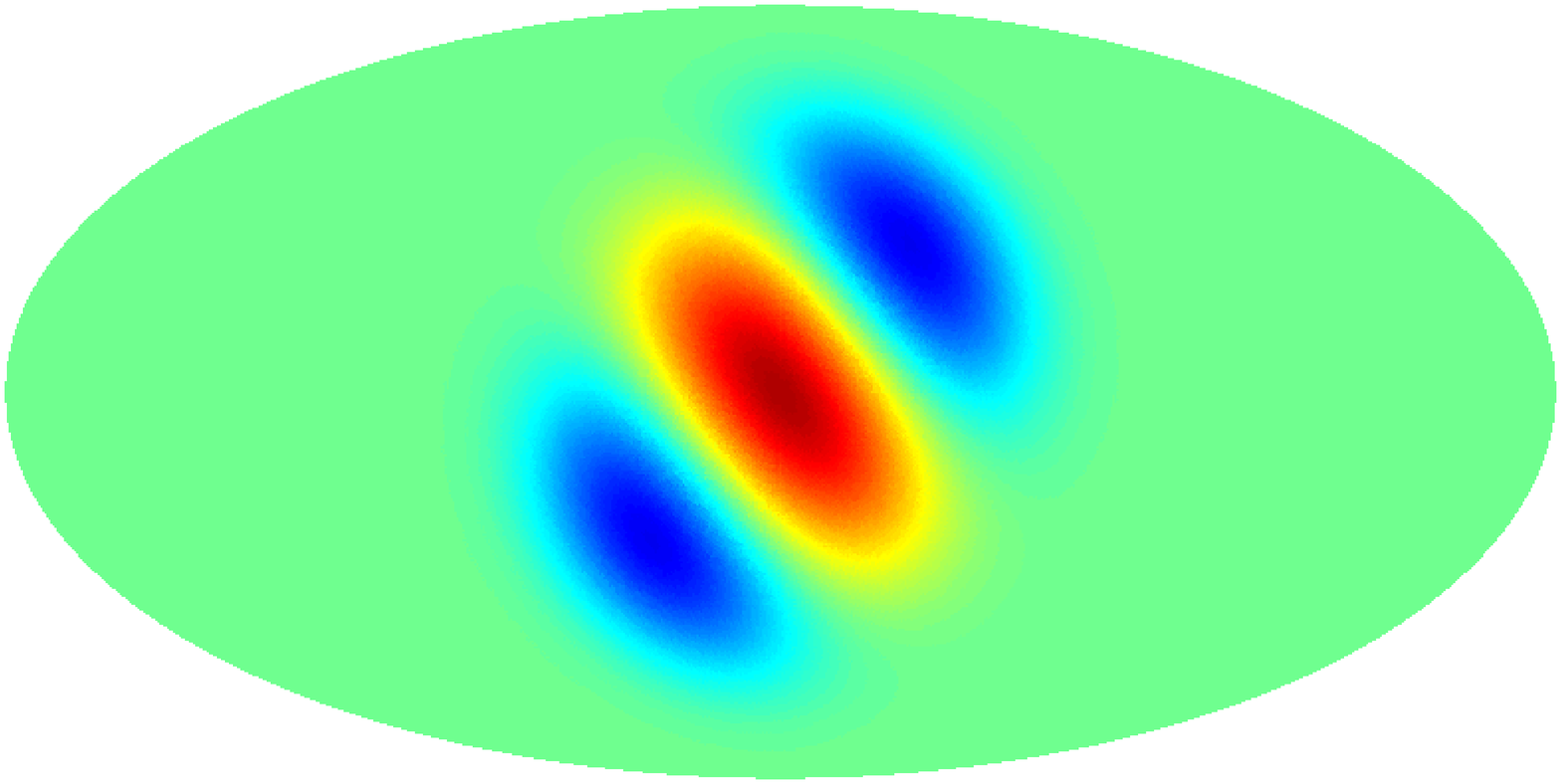}\hspace*{10mm}}
\caption{Mollweide projection of second Gaussian derivative
wavelet on the sphere for wavelet half-width $\halfw\simeq0.4 $
rad.
The rotated wavelet illustrated in panel~(d) can be constructed from a sum of weighted versions of the basis wavelets illustrated in panels~(a) through (c).
}
\label{fig:twodg}
\end{figure}

The derivatives of order $N_{\rm d}$ in direction $\hat{x}$ of radial
functions $\phi(r)$ on the plane are steerable wavelets. The corresponding
steerable wavelets on the sphere produced by inverse stereographic
projection may be rotated in terms of $M=N_{\rm d}+1$ basis filters,
and are band-limited in $\varphi$ at $N=N_{\rm d}+1$. The second derivative
$\Psi^{\partial_{\hat{x}}^{2}}$ of any radial function has a band
limit $N=3$, and contains the frequencies $n=\{0,\pm2\}$ only. $\Psi^{\partial_{\hat{x}}^{2}}$ may
be rotated in terms of three basis filters: the second derivatives
in directions $\hat{x}$ and $\hat{y}$, $\Psi_{1}=\Psi^{\partial_{\hat{x}}^{2}}$
and $\Psi_{2}=\Psi^{\partial_{\hat{y}}^{2}}$, and the cross derivative
$\Psi_{3}=\Psi^{\partial_{\hat{x}}\partial_{\hat{y}}}$, with the
corresponding weights $k_{1}(\chi)=\cos^{2}\chi$, $k_{2}(\chi)=\sin^{2}\chi$,
and $k_{3}(\chi)=\sin2\chi$. \Fig{\ref{fig:twodg}}
illustrates the explicit example of the second derivative of a Gaussian,
which is the wavelet employed in the subsequent analysis. In this case, the original
radial function on the plane is given as $\phi^{(Gauss)}(r)=-\sqrt{4/3\pi}e^{-r^{2}/2}$,
with proper normalisation ensuring that the mother wavelet $\Psi^{\partial_{\hat{x}}^{2}(Gauss)}$
on the sphere is normalised to unity. The explicit basis filters
obtained by differentiation are analytically defined by \citet{wiaux:2005c}.
Recall that the natural dilation on the tangent plane at the North
Pole sends the radial variable $r(\theta)$ onto $r(\theta)/a$. The
value of the scale $a$ therefore identifies with the dispersion of
the Gaussian. Consequently, we define the angular size of our second Gaussian
derivative wavelet on the sphere as twice the \emph{half-width} of the wavelet, where the half-width is defined by 
$\theta_{\rm hw}=2\arctan(a/2)$, which is closely approximated by $a$ at small scales.

%

We use wavelet steerability to define local morphological
measures of real signals $F$ on the sphere. Local geometrical
quantities have previously been defined in real space, and have been
used to analyse \cmb\ data and test cosmological models 
\citep{barreiro:1997,barreiro:2001,monteserin:2005,gurzadyan:2005}.
Our approach in wavelet space is completely novel.  By linearity of the
wavelet decomposition, the steerability
relation also holds on the wavelet coefficients of a signal $F$,
\begin{equation}
W_{\Psi}^{F}\left(\omega_{0},\chi,a\right)=\sum_{m=1}^{M}k_{m}\left(\chi\right)W_{\Psi_{m}}^{F}\left(\omega_{0},a\right) 
\spcend,
\label{eqn:steerablity-coefficient}
\end{equation}
where the coefficients $W_{\Psi_{m}}^{F}(\omega_{0},a)$ result from
the correlation of $F$ with the non-rotated ($\chi=0$) filter $\Psi_{m}$
at scale $a$. Technically these $M$ coefficients thus gather all
the local information retained by the steerable wavelet analysis at
the point $\omega_{0}$ and at the scale $a$. One may be willing
to reorganise this information in terms of $M$ quantities with an
explicit local morphological meaning. For $M=1$, the steerable wavelet
is actually axisymmetric, and only a measure of signed-intensity $\morphi^{F}(\omega_{0},a)$
of local features in the signal is accessible (given by the wavelet coefficient
$W_{\Psi}^{F}(\omega_{0},a)$ directly). For $M\geq2$, the steerable
wavelet is directional and the orientation, or direction,
$\morphc^{F}(\omega_{0},a)$ of local features may be accessed through the value which maximises
the wavelet coefficient. The signed-intensity $\morphi^{F}(\omega_{0},a)$
of the local feature is given by that coefficient of maximum absolute
value. Additional local morphological measures may be defined for $M\geq3$, notably
a measure of the elongation of local features, probing their non-axisymmetry,
in addition to their signed-intensity and orientation. As the wavelet
coefficient in each orientation naturally probes the extension of
the local feature in the corresponding direction, this elongation
$\morphe^{F}(\omega_{0},a)$ can simply be measured from the absolute
value of the ratio of the wavelet coefficient of maximum absolute
value, to the wavelet coefficient in the perpendicular direction.
The higher the number $M$ of basis filters, the wider the accessible
range of local morphological measures.

Any second derivative of a radial function, such as the second Gaussian
derivative, is defined from $M=3$ basis filters. The signed-intensity,
orientation, and elongation are therefore accessible in this case. 
Because the mother wavelet
$\Psi^{\partial_{\hat{x}}^{2}}$ oscillates in the tangent direction
$\hat{x}$, it actually detects features oriented along the tangent
direction $\hat{y}$. The orientation of local features is therefore
given as 
\begin{equation}
\label{eqn:morphc}
\frac{\pi}{2}\leq\quad
\morphc^{F}\left(\omega_{0},a\right)
\equiv
\chi_{0}+\frac{\pi}{2}\quad<\frac{3\pi}{2}
\spcend,
\end{equation}
where the orientation $\chi_{0}\equiv\chi_{0}(\omega_{0},a)$ that
maximises the wavelet coefficient can be analytically defined from
\eqn{\ref{eqn:steerablity-coefficient}}. 
Notice that the second derivative of a radial function
is invariant under a rotation by $\pi$ around itself. Consequently,
the local orientations define headless vectors in the tangent plane
at $\omega_{0}$. By convention the corresponding vectors are chosen
to point towards the Northern hemisphere only, and the local orientations
are therefore in the range $[\pi/2,3\pi/2)$.
For a second derivative of a radial function, as for any real steerable
wavelet, the signed-intensity of local features is given by the wavelet coefficient
in the direction $\chi_{0}$:
\begin{equation}
\label{eqn:morphi}
\morphi^{F}\left(\omega_{0},a\right)
\equiv
W_{\Psi^{\partial_{\hat{x}}^{2}}}^{F}\left(\omega_{0},\chi_{0},a\right)
\spcend .
\end{equation}
The elongation of local features analysed by the second derivative
of a radial function is explicitly defined by
\begin{equation}
0\leq\quad \morphe^{F}\left(\omega_{0},a\right)
\equiv
1-\Biggl\vert\frac{W_{\Psi^{\partial_{\hat{x}}^{2}}}^{F}\left(\omega_{0},\chi_{0}+\frac{\pi}{2},a\right)}{W_{\Psi^{\partial_{\hat{x}}^{2}}}^{F}\left(\omega_{0},\chi_{0},a\right)}
\Biggr\vert\quad\leq1
\spcend.\label{elongation2GD}
\end{equation}
Numerical tests performed on elliptical Gaussian-profile features
show that this elongation measure for the second
Gaussian derivative increases monotonously in the range $[0,1]$ with
the intrinsic eccentricity $e\in[0,1]$ of the features. 
While it is possible to define alternative elongation measures, these numerical tests indicate that the chosen definition is not an arbitrary measure of the non-axisymmetry of local features, but represents a rough estimate of the eccentricity of
a Gaussian-profile local feature.

\section{Analysis procedures}
\label{sec:procedures}

The data and analysis procedures that we use in an attempt to detect the \isw\ effect are described in detail in this section.  We propose two distinct analysis techniques based on the correlation of the local morphological measures on the sphere defined in \sectn{\ref{sec:steerability}}.  After describing the data, we discuss the generic procedure to search for any correlation between the \nvss\ and \wmap\ data.  We then describe in detail the two approaches proposed, motivating the analyses and highlighting the differences between them.

\subsection{Data and simulations}
\label{sec:data}

In this work we examine the three-year release of the \wmap\ data and the \nvss\ data for correlations induced by the \isw\ effect.  Here we briefly describe the data, preprocessing of the data and the simulations performed to quantify any correlations detected in the data.

We examine the co-added three-year \wmap\ sky map.  This map is constructed from a noise weighted sum of the maps observed by the Q-, V- and W- channels of \wmap, in order to enhance the signal-to-noise ratio of the resultant map.  The co-added map was first introduced by the \wmap\ team in their non-Gaussianity analysis \citep{komatsu:2003} and has since been used in numerous analyses.  We use the template based foreground cleaned \wmap\ band maps \citep{bennett:2003b} to construct the co-added map.  The conservative \kpzero\ mask provided by the \wmap\ team is used to remove remaining Galactic emission and bright point sources.
Since the \isw\ effect is expected to induce correlations on scales $>2^\circ$ \citep{afshordi:2004}, we downsample the co-added map to a pixel size of $\sim\!\!55\arcmin$ (a \healpix\footnote{\url{http://healpix.jpl.nasa.gov/}} \citep{gorski:2005} resolution of $\nside=64$).  This reduces the computation requirements of the subsequent analysis considerably, while ensuring the analysis remains sensitive to \isw\ induced correlations.

The large sky coverage and source distribution  of the \nvss\ data make it an ideal probe of the local matter distribution to use when searching for the \isw\ effect.  Sources in the catalogue are thought to be distributed in the range $0<z<2$, with a peak distribution at $z\sim0.8$ \citep{boughn:2002}.  This corresponds closely to redshift regions where the \isw\ signal that we hope to detect is produced \citep{afshordi:2004}.  The \nvss\ source distribution is projected onto the sky in a \healpix\ representation at the same resolution as the \wmap\ co-added map considered (\ie\ $\nside=64$) and an important systematic in the data is corrected.  The correction of this systematic and the resulting preprocessed \nvss\ data examined here are both identical to that analysed in \citet{mcewen:2006:isw}, hence we refer the interested reader to this work for more details of the data preprocessing.  Not all of the sky is sufficiently observed in the \nvss\ catalogue.  We construct a joint mask to exclude from our subsequent analysis those regions of the sky not observed in the catalogue, in addition to the regions excluded by the \wmap\ \kpzero\ mask.  The \wmap\ and \nvss\ data analysed subsequently, with the joint mask applied, are illustrated in \fig{\ref{fig:isw_maps}}.

In order to constrain the statistical significance of any detection of correlation between the \wmap\ and \nvss\ data we perform Monte Carlo simulations.  1000 simulations of the \wmap\ co-added map are constructed, modelling carefully the beam and anisotropic noise properties of each of the \wmap\ channels and mimicking the co-added map construction procedure (including the downsampling stage).  The analyses subsequently performed on the data are repeated on the simulations and compared.  

\newlength{\mapplotwidth}
\setlength{\mapplotwidth}{75mm}

\begin{figure}
\centering
\subfigure[\wmap]{\includegraphics[clip=,width=\mapplotwidth]{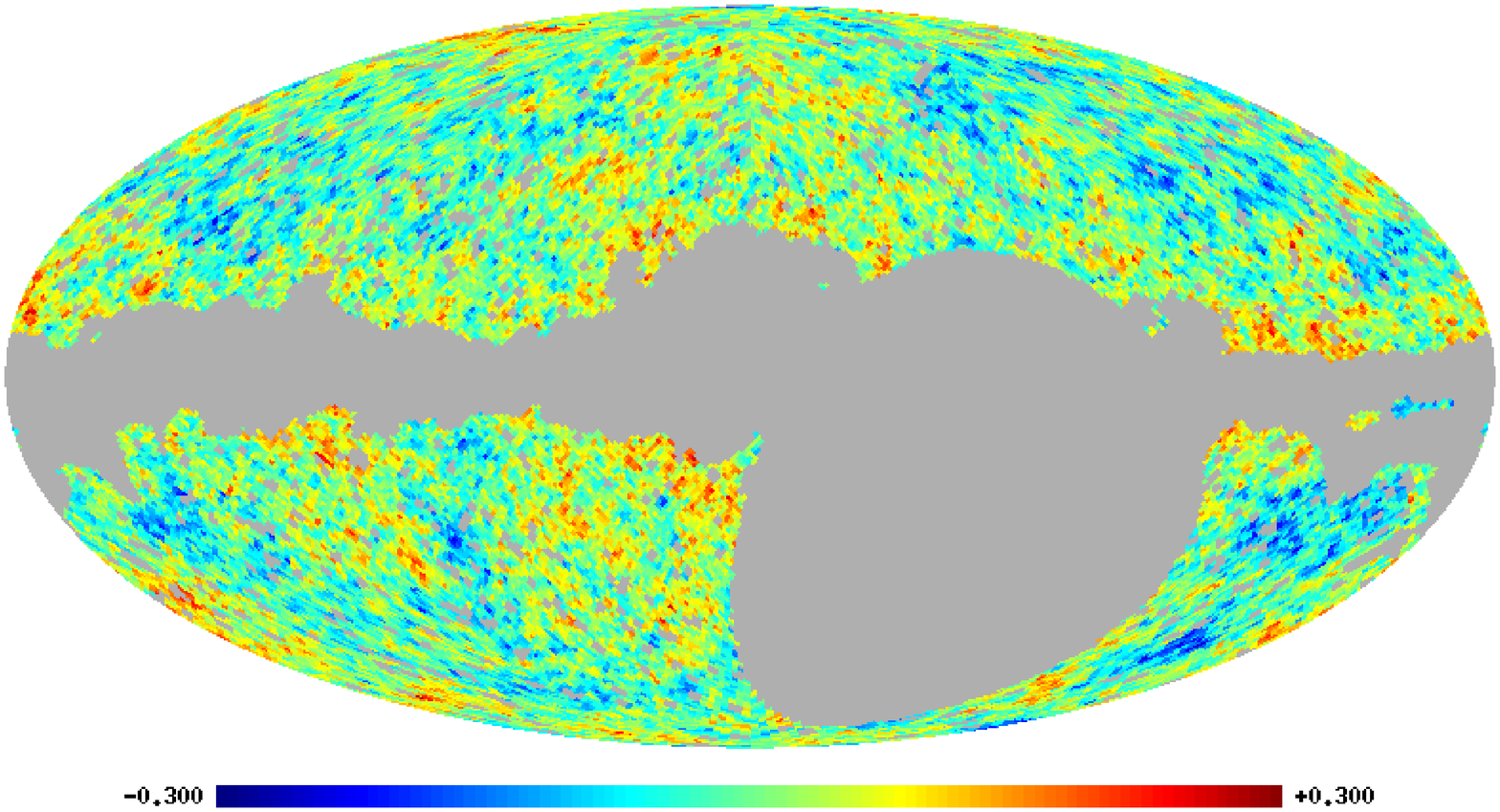}}
\subfigure[\nvss]{\includegraphics[clip=,width=\mapplotwidth]{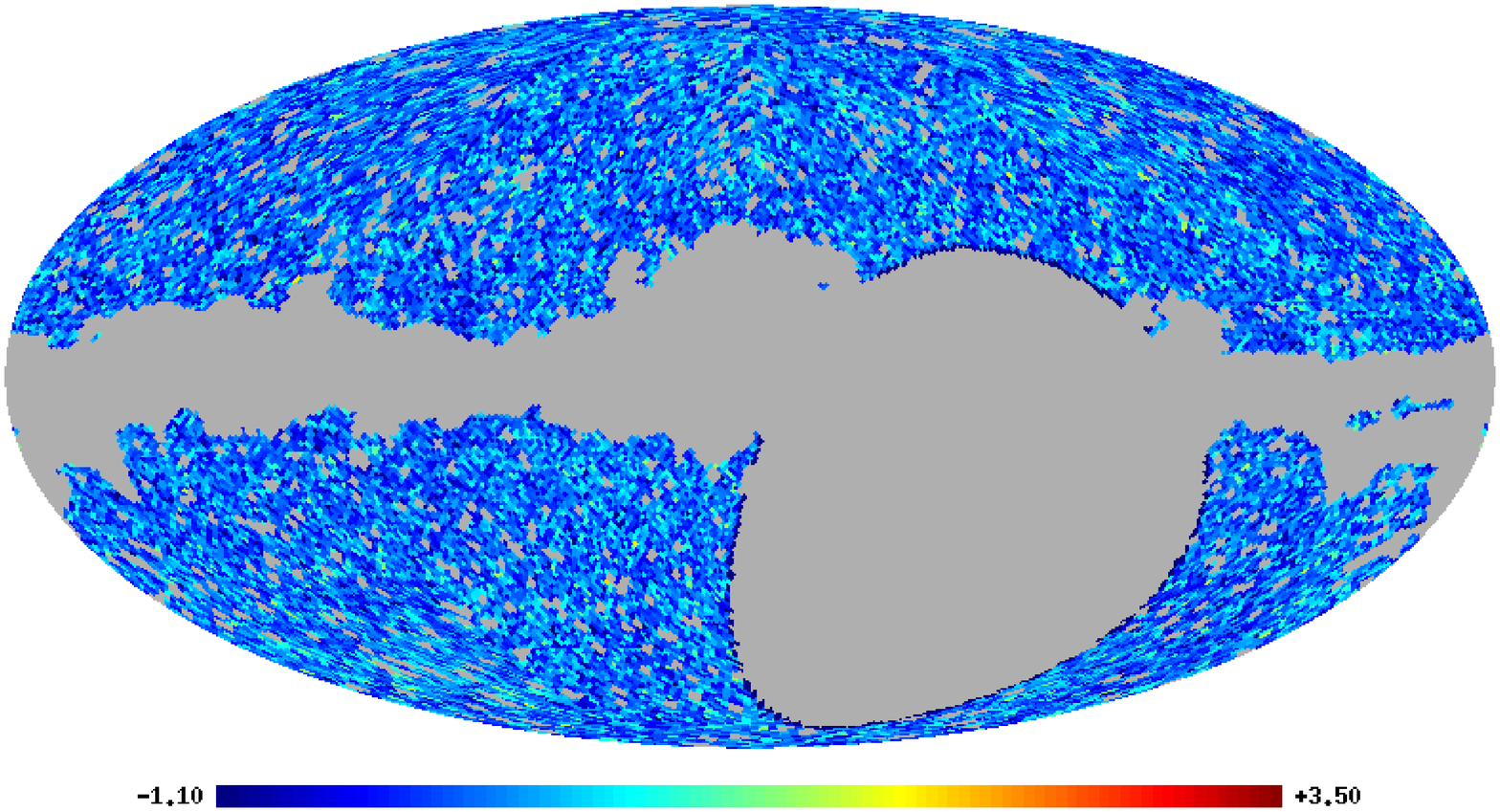}}
\caption{\wmap\ co-added three-year and \nvss\ maps (Mollweide projection) after application of the joint mask.  The maps are downsampled to a pixel size of $\sim\!\!55\arcmin$.  The \wmap\ temperature data are reported in mK, while the \nvss\ data are reported in number-of-counts per pixel.}
\label{fig:isw_maps}
\end{figure}

\subsection{Generic procedure}
\label{sec:procedure_generic}

Two different analysis procedures are applied to test the data for correlation.  The first procedure correlates the local morphological measures of signed-intensity, orientation and elongation, extracted independently from the two data sets. This procedure does not rely on any assumption about the correlation in the data.  The second procedure correlates local features in the \wmap\ data that are matched in orientation to local features extracted from the \nvss\ data.  This approach explicitly assumes that local features of the \lss\ are somehow included in the \cmb.
%
%
%
In the absence of any \isw\ effect one would not expect to detect any significant correlation in either of these analyses.  The analysis procedures are described in more detail in the following subsections.  Firstly, we describe the generic part of the procedure common to both techniques.

We consider only those scales where the \isw\ effect is expected to be significant \citep{afshordi:2004}, \ie\ scales approximately corresponding to wavelet half-widths $\theta_{\rm hw}$ of 
$\{100\arcmin, 150\arcmin, 200\arcmin, 250\arcmin, 300\arcmin, 400\arcmin, 500\arcmin, 600\arcmin\}$.  The local morphological measures defined in \sectn{\ref{sec:steerability}} are computed for the \wmap\ and \nvss\ data and are used to compute various correlation statistics (see the following subsections for more detail).  Identical statistics are computed for the Monte Carlo simulations in order to measure the statistical significance of any correlation detected in the data.  A statistically significant correlation in the data apparent from any of these statistics is an indication of the \isw\ effect, provided it is not due to foreground contributions or systematics.

The procedure described previously would be sufficient for data with full-sky coverage.  Unfortunately this is not the case and the application of the joint mask must be taken into account.  The application of the mask distorts those morphological quantities constructed from wavelets with support that overlaps the mask exclusion regions.  The associated local morphological measures must be excluded from the analysis.  An extended mask is computed for each scale to remove all contaminated values.  The extended masks are constructed by extending the central masked region by the wavelet half-width, whilst maintaining the size of point source regions in the mask.  We use identical masks to those applied by \citet{mcewen:2006:isw} and refer the interested reader to this work for more details.

Before proceeding with our morphological analyses it is important to check that the \wmap\ and \nvss\ data do not contain predominantly axisymmetric features, corresponding to local elongation trivially equal to zero and for which no local orientation might be defined.
To do this we examine the distribution of our morphological measure of elongation defined in \sectn{\ref{sec:steerability}}.  
In \fig{\ref{fig:elong_hist}} we plot the histograms of the elongation measures computed from the data (outside of the extended exclusion masks).  These histograms are built from all scales but the distributions obtained for each individual scale also exhibit similar structure.  Many elongation values computed from both data sets lie far from zero, thereby justifying the morphological analyses that we propose. 

In order to give some intuition on the magnitude of various statistics computed in the remainder of this paper, we also quote the orders of magnitude for the means and standard deviations of the different local morphological measures, as estimated from the two data sets.  Firstly, the mean and standard deviation of the elongation in the \wmap\ data are respectively $\hat{\mu}^\tplab_\morphe=0.61$ and $\hat{\sigma}^\tplab_\morphe=0.27$, when data from all scales are gathered. The corresponding values for the \nvss\ data are $\hat{\mu}^\ndlab_\morphe=0.62$ and $\hat{\sigma}^\ndlab_\morphe=0.26$. Secondly, recall that the galaxy density and \cmb\ random fields on the sphere are assumed to be homogeneous and isotropic. The local orientations computed from the data sets should therefore reflect a uniform distribution in the range $[\pi/2,3\pi/2)$. Consistently, the mean and standard deviation of the orientation in the WMAP data are respectively $\hat{\mu}^\tplab_\morphc=3.10$ and $\hat{\sigma}^\tplab_\morphc=0.93$, again gathering data from all scales. The corresponding values for the NVSS data are $\hat{\mu}^\ndlab_\morphc=3.10$ and $\hat{\sigma}^\ndlab_\morphc=0.90$. Finally, notice that the local signed-intensity of a signal is expressed, \emph{a priori} on the whole real line, in the same units as the signal itself. The signed-intensity of the \wmap\ temperature fluctuation data is given in mK, while the signed-intensity of the \nvss\ galaxy source count fluctuation data is given in number-of-counts per pixel (see \fig{\ref{fig:isw_maps}}). The mean and standard deviation of the signed-intensity in the \wmap\ data are respectively $\hat{\mu}^\tplab_\morphi=-2\times10^{-4}$ and $\hat{\sigma}^\tplab_\morphi=6\times10^{-3}$, still accounting for data from all scales. The corresponding values for the \nvss\ data are $\hat{\mu}^\ndlab_\morphi=0.001$ and $\hat{\sigma}^\ndlab_\morphi=0.01$.

\newlength{\histplotwidth}
\setlength{\histplotwidth}{60mm}

\begin{figure}
\centering
\subfigure[\wmap]{\includegraphics[clip=,width=\histplotwidth]{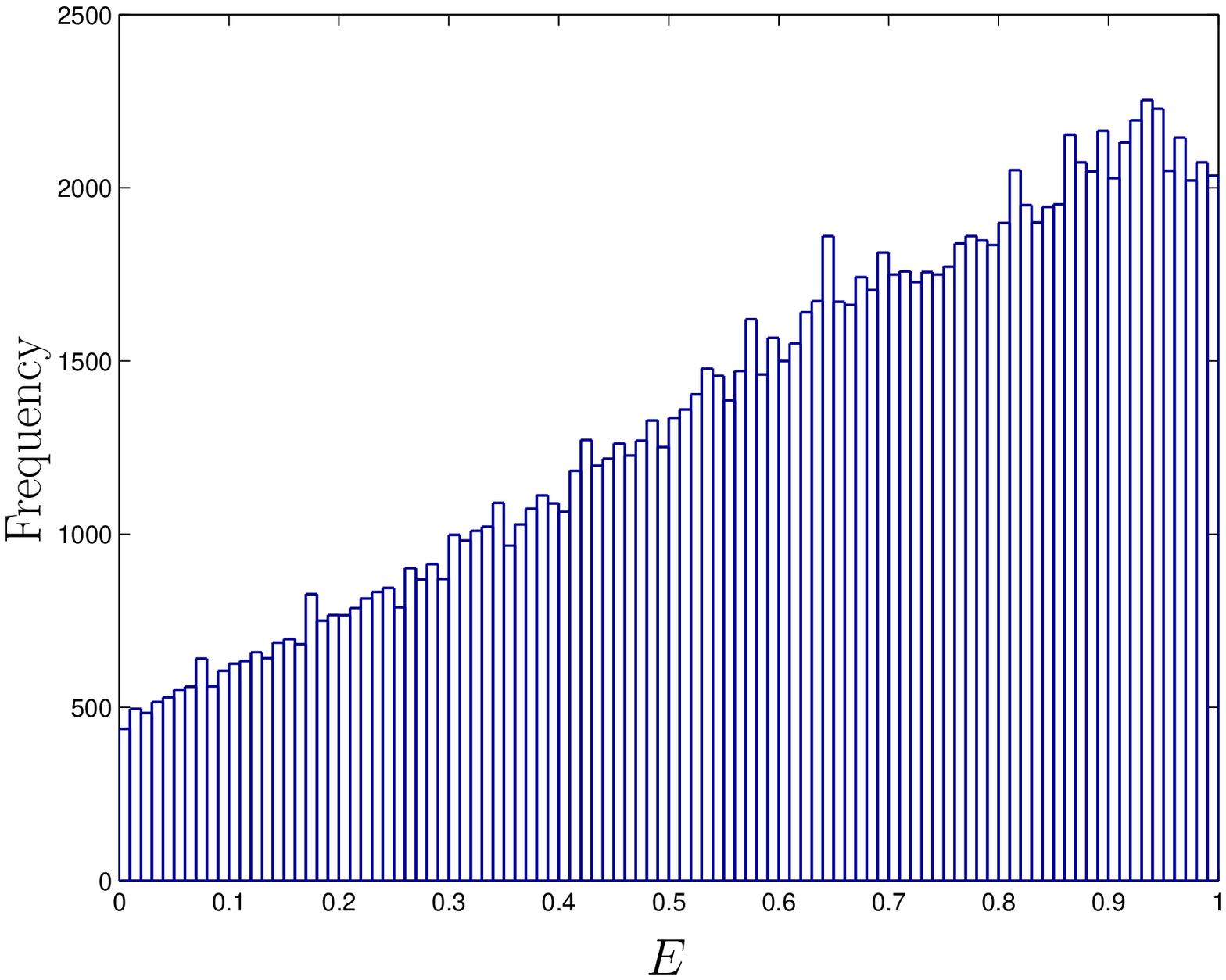}} 
\subfigure[\nvss]{\includegraphics[clip=,width=\histplotwidth]{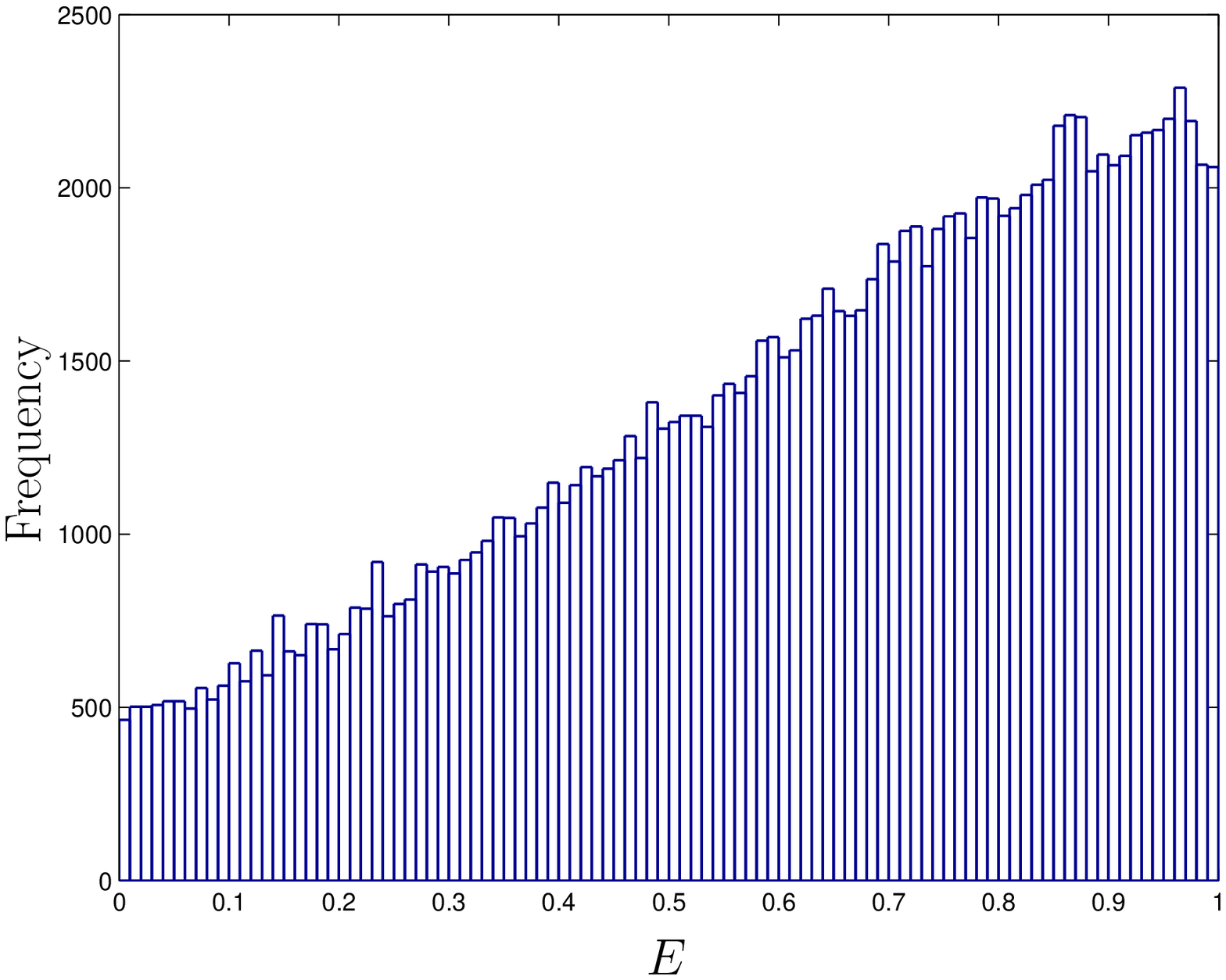}}
\caption{Histograms of the elongation values computed from the \wmap\ and \nvss\ data for all scales.  Many elongation values lie far from zero, thereby justifying the morphological analyses that we propose.}
\label{fig:elong_hist}
\end{figure}

\subsection{Local morphological analysis}
\label{sec:procedure_morph}

In the local morphological analysis that follows we correlate the \wmap\ and \nvss\ data through each of the morphological measures provided by the \twogd\ wavelet separately, \ie\ the signed-intensity, orientation and elongation of local features.  Local features are extracted independently from the \wmap\ and \nvss\ data.  We then compute the correlation by
\begin{eqnarray}
X_{\morphs_i}^{\ndlab \tplab}(\scalea) & = &\frac{1}{N_{\rm p}} \sum_{\sa_0} \morphs_i^\ndlab(\sa_0, a) \: \morphs_i^\tplab(\sa_0, a)  \nonumber \\
& & - \biggl[ \frac{1}{N_{\rm p}} \sum_{\sa_0} \morphs_i^\ndlab(\sa_0, a) \biggr] \cdot \biggl[ \frac{1}{N_{\rm p}} \sum_{\sa_0} \morphs_i^\tplab(\sa_0, a) \biggr]  \spcend ,
\label{eqn:covest}
\end{eqnarray}
where $\morphs_i = \{
\morphi,\:
\morphc,\:
\morphe 
\}$
is the morphological measure examined, with superscript \ndlab\ or \tplab\ representing respectively the \nvss\ and \wmap\ data (note that the morphological measures are computed at  
$N_{\rm p}$ discrete points on the sky only).  This analysis provides the most general approach to computing correlations and does not make any assumption about possible correlation in the data.  In the absence of an \isw\ effect the \wmap\ and \nvss\ data should be independent and none of the correlation estimators defined by \eqn{\ref{eqn:covest}} should exhibit a significant deviation from zero.

\subsection{Matched intensity analysis}
\label{sec:procedure_intensity}

In the matched intensity analysis that follows we first compute the orientation and signed-intensity of local features for the \nvss\ data.  Using the local orientations extracted from the \nvss\ data, we compute the signed-intensity of local features in the \wmap\ data that are matched in orientation to the local features in \nvss\ data.
For this matched intensity analysis it no longer makes sense to examine the correlation of orientations any further since they are identical in both data sets, nor to examine the elongation measures any further since the elongation does not measure the eccentricity of local features in the \wmap\ data when the orientation is extracted from the \nvss\ data.  We therefore only correlate, through \eqn{\ref{eqn:covest}} again, the signed-intensity of features extracted from the data.  This analysis is based on the assumption that local features in the \nvss\ data are somehow included in the \wmap\ data.  
This is a conceptually different analysis to the local morphological analysis described in the preceding subsection.  

\begin{figure}
\centering
\includegraphics[clip=,width=\histplotwidth]{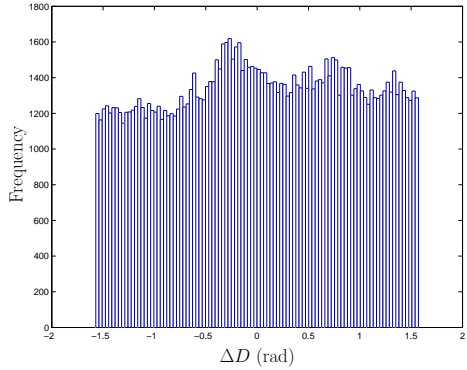}
\caption{Histogram of the difference in orientation of local features extracted independently in the \wmap\ and \nvss\ data for all scales.  
This difference $\Delta \morphc$ is defined modulo $\pi$, in the range $[-\pi/2,\pi/2)$.
The distribution is not concentrated about zero, indicating that the local morphological and matched intensity analyses are indeed different practically, as well as conceptually.}
\label{fig:delta_chi_hist}
\end{figure}

Before proceeding with this analysis it is important to check that it differs in practice to the local morphological analysis.  To do this we examine the difference in the orientation of local features, extracted independently in each data set, at each position on the sky.  From relation \eqn{\ref{eqn:morphc}} it can be seen that this difference technically lies in the range $[-\pi,\pi)$. It is however defined modulo $\pi$, and the range is explicitly restricted to $\Delta \morphc \in [-\pi/2,\pi/2)$.
If orientation differences are predominantly zero, then we know \emph{a priori} that the two analysis procedures are essentially the same and would give the same result for the correlation of signed-intensities.  In \fig{\ref{fig:delta_chi_hist}} we plot the histogram of the differences in orientation between the two data sets (outside of the extended exclusion masks), built from all scales (the distributions obtained for each individual scale exhibit similar structure).  The orientation differences are broadly distributed about zero, thus the two analysis techniques proposed are indeed different practically, as well as conceptually.
Moreover, the relative flatness of the distribution suggests no obvious correlation of orientations between the \wmap\ and \nvss\ data. 
This is examined in detail through the local morphological analysis when applied to orientations.  

Notice that the local morphological measure of the orientation defined by \eqn{\ref{eqn:morphc}} depends on the coordinate system considered through the definition of the origin $\chi=0$ at each point on the sphere.  The estimator for the correlation of orientations between the two data sets given by \eqn{\ref{eqn:covest}} also depends on the coordinate system. This dependence does not affect the statistical significance of any correlation detected, which simply measures a possible anomaly between the data and the simulations.
One could substitute the estimator \eqn{\ref{eqn:covest}} by a measure of flatness for the distribution of the difference in the orientation $\Delta \morphc$ between the two data sets. This difference is indeed independent of the coordinate system and has a direct physical meaning. However, this would simply provide an alternative measure of the statistical significance of any correlation detected in the orientation of local features.  Either estimator is therefore suitable; for consistence we choose the former case.



\section{Local morphological correlations}
\label{sec:morph}

In this section we present the results obtained from the local morphological analysis described in \sectn{\ref{sec:procedure_morph}} to examine the \wmap\ and \nvss\ data for possible correlation.  Correlations are detected and are examined to determine whether they are due to foreground contamination or instrumental systematics, or whether they are indeed induced by the \isw\ effect.

\subsection{Detections}
\label{sec:morph_detections}

The correlation statistics computed from the \wmap\ and \nvss\ data for each morphological measure are displayed in \fig{\ref{fig:stat_morph}}.  Significance levels computed from the 1000 Monte Carlo simulations described in \sectn{\ref{sec:data}} are also shown on each plot.  
A non-zero correlation signal is clearly present in the signed-intensity of local features.
A strong detection at 99.3\% significance is made in the correlation of signed-intensities for wavelet half-width $\halfw\simeq400\arcmin$.  This correlation deviates from the mean of the Monte Carlo simulations by 2.8 standard deviations.
Moderate detections are also made in the correlation of the orientations and elongations of local features for $\halfw\simeq400\arcmin$ at 93.3\% significance and for $\halfw\simeq600\arcmin$ at 97.2\% significance respectively.

The statistical significance of these findings is based on an \emph{a posteriori} selection of the scale corresponding to the most significant correlation.  Nevertheless, such statistics have been widely used to quantify detections of the \isw\ effect, and are satisfactory provided that the \emph{a posteriori} nature of the analysis is acknowledged.  However, in this work we also consider a \chisqd\ analysis to examine the significance of correlation between the data when the correlation statistics for all scales are considered in aggregate.  This \emph{a priori} statistic is more conservative in nature; we consider it as a simple alternative detection measure which takes into account the correlation between scales.  For the correlation computed for each local morphological measure we compute the \chisqd\ defined by 
\begin{equation}
\chisqd_{\morphs_i} =  
{\vect{X}_{\morphs_i}^{\ndlab \tplab}} ^\dagger \:
{\rm C}_{\morphs_i}^{-1} \:
\vect{X}_{\morphs_i}^{\ndlab \tplab}
\spcend ,
\end{equation}
where $\vect{X}_{\morphs_i}^{\ndlab \tplab}$ is a concatenated vector of $X_{\morphs_i}^{\ndlab \tplab}(\scalea)$ for all scales, the matrix ${\rm C}_{\morphs_i}$ is a similarly ordered covariance matrix of the morphological correlation statistics computed from the Monte Carlo simulations and ${}^\dagger$ represents the matrix transpose operation.  Comparing the \chisqd\ values computed from the data to those computed from the Monte Carlo simulations, it is possible to quantify the statistical significance of correlation in the data.  
Moderate detections are made in the correlation of signed-intensities and orientations of local features at 94.5\% and 93.2\% significance respectively.  
The detection in the correlation of elongations drops considerably to 84.5\% significance.
%

\begin{figure}
\centering
\subfigure[\wmap\ for independent features]{\includegraphics[clip=,width=\mapplotwidth]{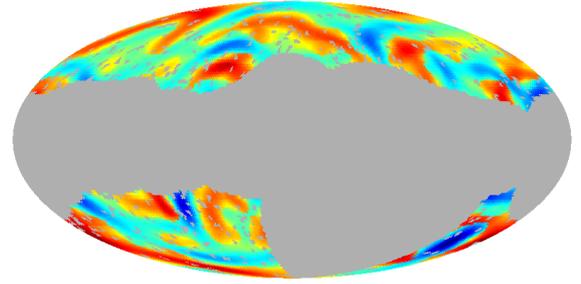}}
\subfigure[\wmap\ for features  matched to \nvss\ orientations]{\includegraphics[clip=,width=\mapplotwidth]{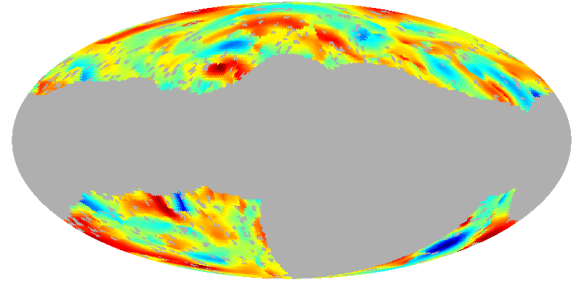}}
\subfigure[\nvss]{\includegraphics[clip=,width=\mapplotwidth]{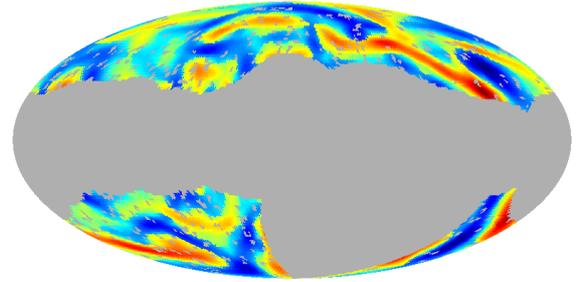}}
\caption{Morphological signed-intensity maps (Mollweide projection) corresponding to the scale \mbox{($\halfw\simeq400\arcmin$)} on which the maximum detections of correlation are made.  In panel~(a) signed-intensities are shown for local features extracted independently from the \wmap\ co-added data, whereas in panel~(b) signed-intensities are shown for local features in the \wmap\ co-added data that are matched in orientation to local features in the \nvss\ data.  Due to the strength of the correlation in the data, it is possible to observe the correlation both between maps (a) and (c) and between maps (b) and (c) by eye.
}
\label{fig:wcoeff_maps}
\end{figure}

The steerable wavelet analysis performed allows one to localise on the sky those regions that contribute most significantly to the correlation detected in the data.  This type of localisation was performed by \citet{mcewen:2006:isw}.  However, it was found in this work that, although the localised regions were a significant source of correlation between the data, they by no means provided the sole contribution of correlation.  Similar findings were also obtained by \citet{boughn:2005} using a different analysis technique.  These results are consistent with intuitive predictions based on the \isw\ effect, namely that one would expect to observe correlations weakly distributed over the entire sky, rather than only a few highly correlated localised regions.  For these reasons we do not repeat the localisation performed by \citet{mcewen:2006:isw} for the current analysis (especially since we shown in \sectn{\ref{sec:morph_foregrounds}} that the detection is likely due to the \isw\ effect and not contamination or systematics).  Instead, we display in \fig{\ref{fig:wcoeff_maps}} the signed-intensity values on the sky for the scale ($\halfw\simeq400\arcmin$) corresponding to the most significance detection of correlation.  Due to the strength of the correlation, it is possible to see it by eye between the \wmap\ (\fig{\ref{fig:wcoeff_maps}}~(a)) and \nvss\ data (\fig{\ref{fig:wcoeff_maps}}~(c)).  We consider only the signed-intensity of local features here since this morphological measure corresponds to the most significant detection of correlation made in these local morphological analyses.

\newlength{\statplotwidth}
\setlength{\statplotwidth}{55mm}

\begin{figure}
\centering
\subfigure[Signed-intensity]{
\includegraphics[width=\statplotwidth]{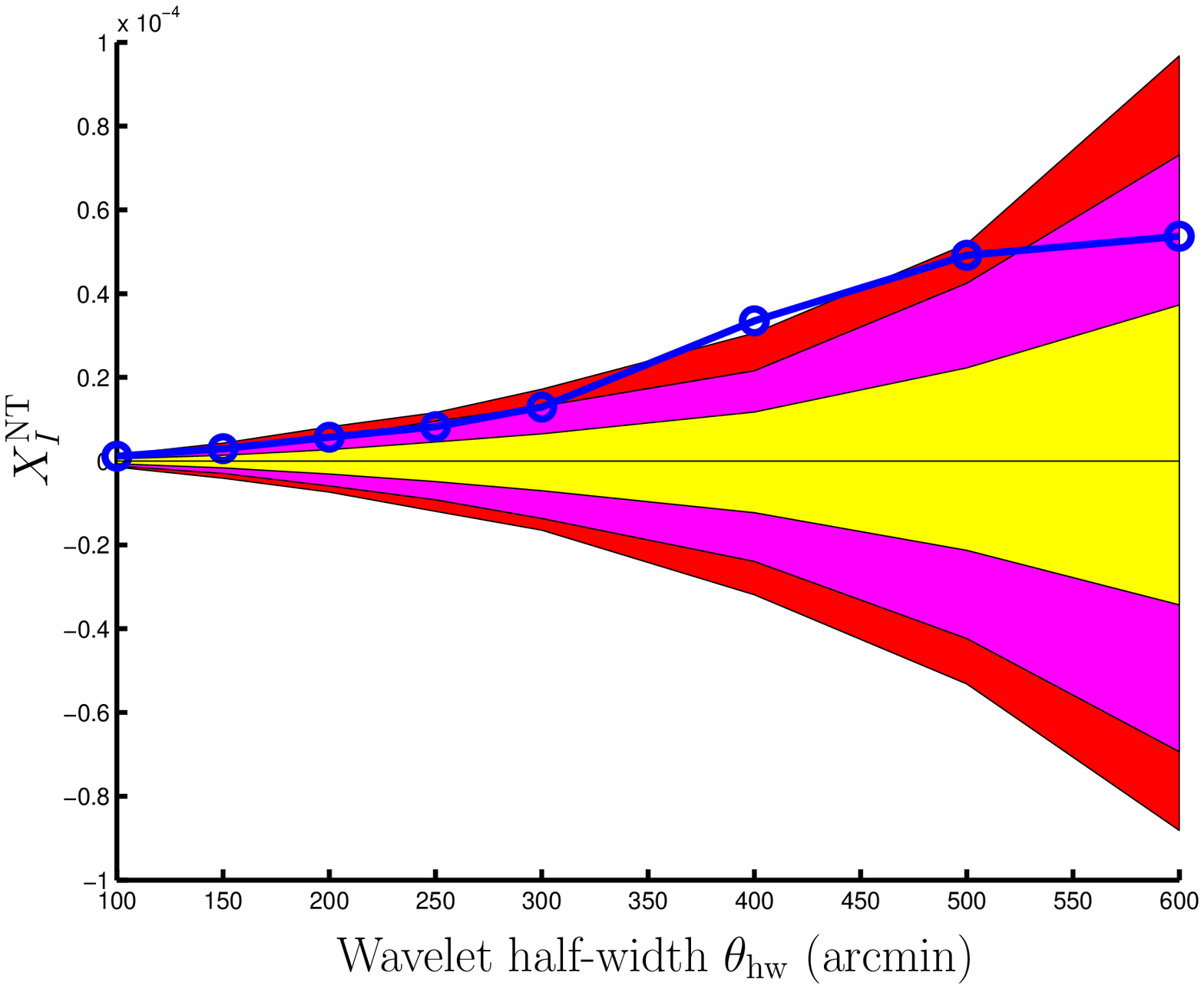}
}\\
\subfigure[Orientation]{
\includegraphics[width=\statplotwidth]{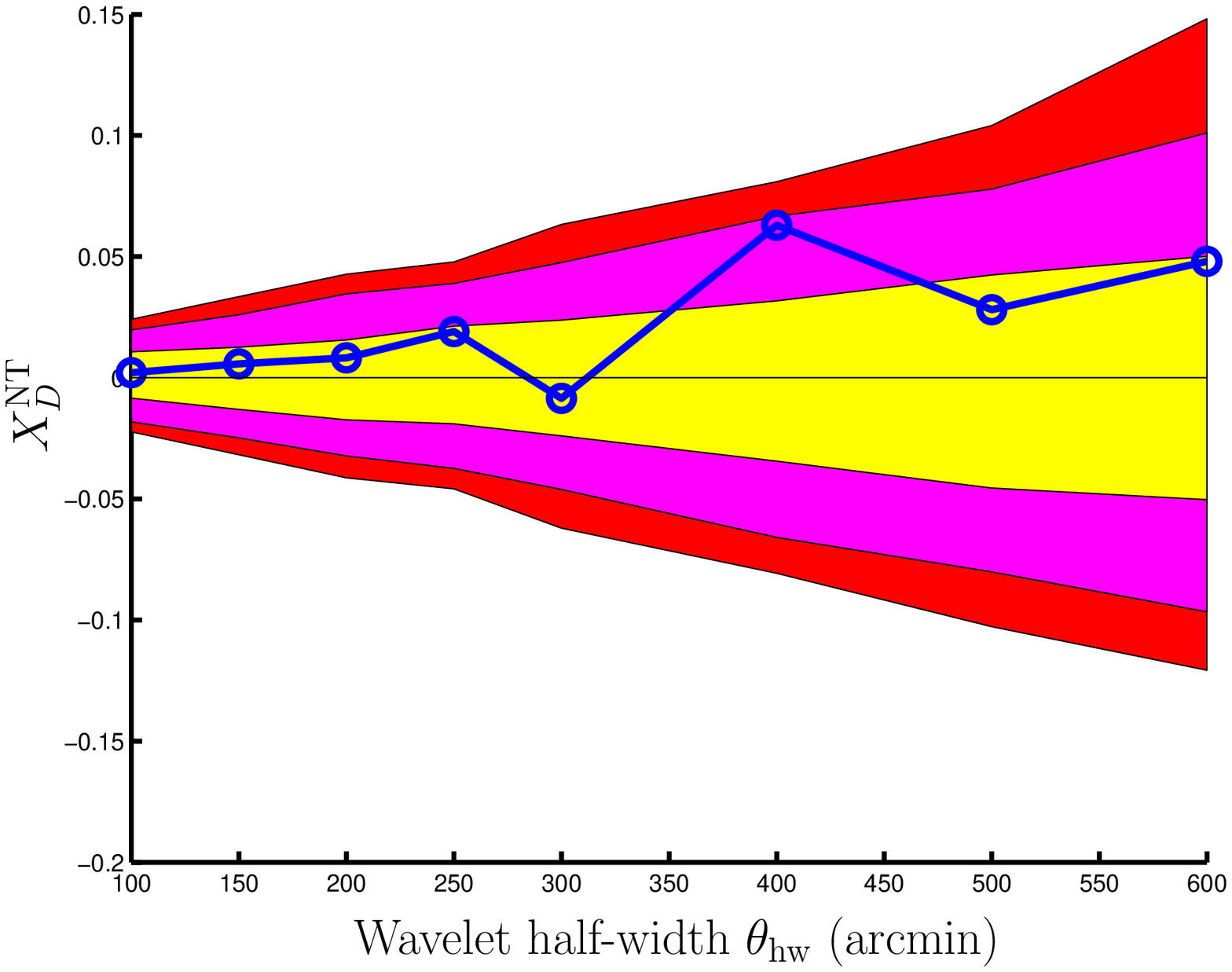}
}\\
\subfigure[Elongation]{
\includegraphics[width=\statplotwidth]{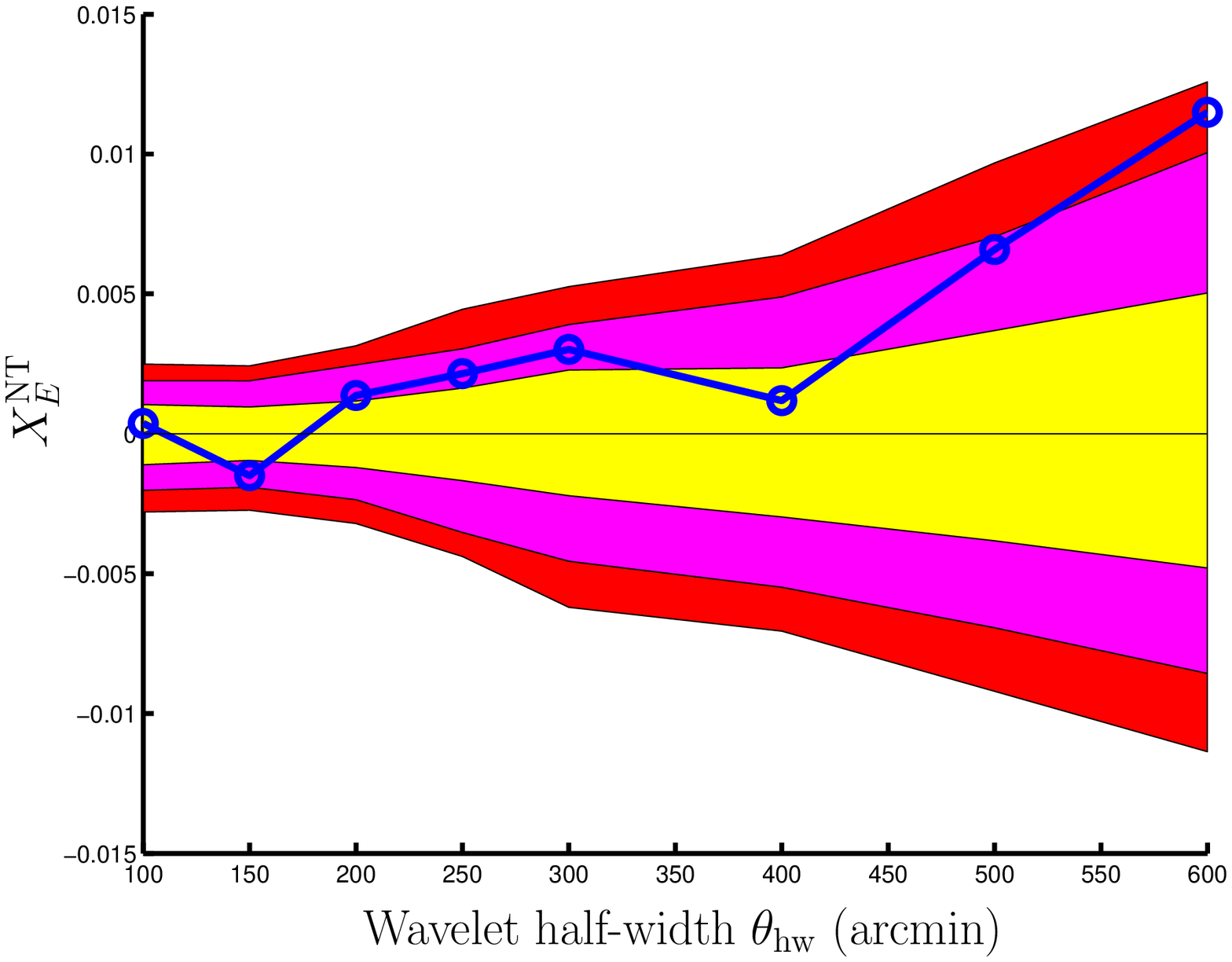}
}
\caption{Correlation statistics computed for each morphological measure in the local morphological analysis, from the \wmap\ co-added map and the \nvss\ map. Significance levels obtained from the 1000 Monte Carlo simulations are shown by the shaded regions for 68\% (yellow/light-grey), 95\% (magenta/grey) and 99\% (red/dark-grey) levels.
The different ranges of the correlations for the signed-intensity, orientation, and elongation are consistent with our rough estimations of the standard deviations for each local morphological measure in each of the two data sets (see \sectn{\ref{sec:procedure_generic}}).
}
\label{fig:stat_morph}
\end{figure}

\subsection{Foregrounds and systematics}
\label{sec:morph_foregrounds}

\begin{figure}
\centering
\subfigure[Individual band maps]{
\includegraphics[width=\statplotwidth]{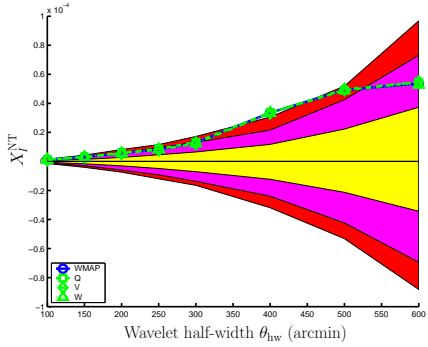}
}\\
\subfigure[Difference maps]{
\includegraphics[width=\statplotwidth]{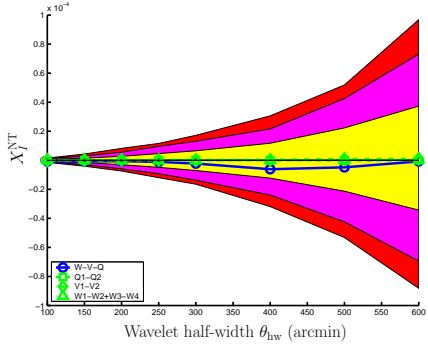}
}
\caption{Correlation statistics computed for the signed-intensity of features in the local morphological analysis, from the \wmap\ individual band and difference maps and the \nvss\ map. Significance levels obtained from the 1000 Monte Carlo simulations are shown by the shaded regions for 68\% (yellow/light-grey), 95\% (magenta/grey) and 99\% (red/dark-grey) levels.}
\label{fig:stat_morph_test}
\end{figure}

To test whether the correlation detected in the preceding subsection is perhaps due to foreground contamination or instrumental systematics in the \wmap\ data, we examine the correlation signals computed using individual \wmap\ bands and difference maps in place of the \wmap\ co-added map.  Again, we consider only the signed-intensity of local features here since this morphological measure corresponds to the most significant detection of correlation made in these local morphological analyses.  In \fig{\ref{fig:stat_morph_test}~(a)} we plot the morphological signed-intensity correlation of the \nvss\ data with each of the individual \wmap\ band maps (constructed from the mean signal observed by all receivers in that band), while in \fig{\ref{fig:stat_morph_test}~(b)} we plot the correlation for maps constructed from band differences (constructed from differences of signals observed by various receivers).

Any correlation induced by unremoved foreground contamination in the \wmap\ data is expected to exhibit a frequency dependence, reflecting the emission law of the foreground component.  However, the same correlation signal is observed in each of the \wmap\ bands and no frequency dependence is apparent (see \fig{\ref{fig:stat_morph_test}~(a)}).  Furthermore, the difference map W$-$V$-$Q is clearly contaminated by foreground contributions, while having a minimal \cmb\ contribution.  No correlation is detected in this difference map (see \fig{\ref{fig:stat_morph_test}~(b)}).  Consequently, we may conclude that foreground contamination is not the source of the correlation detected.

The same correlation signal is observed in each \wmap\ band (see \fig{\ref{fig:stat_morph_test}~(a)}).  Furthermore, this signal is not present in any of the  difference maps constructed for each given band (see \fig{\ref{fig:stat_morph_test}~(b)}), which contain no \cmb\ or foreground contributions.  These findings suggest that the correlation detected is not due to systematics in the \wmap\ data.

The preliminary tests performed in this subsection indicate that it is unlikely that either foregrounds or systematics are responsible for the correlation detected in the signed-intensity of local features. This strongly suggests that the \isw\ effect is responsible for the statistically significant correlation detected in this section between the \wmap\ and \nvss\ data.

\section{Matched intensity correlation}
\label{sec:intensity}

In this section we present the results obtained from the matched intensity analysis described in \sectn{\ref{sec:procedure_intensity}} to examine the \wmap\ and \nvss\ data for possible correlation.  In this setting the orientation of local features extracted from the \wmap\ data at each position on the sky are matched to orientations extracted from the \nvss\ data.  It therefore no longer makes sense to correlate the orientation or elongation of features between the two data sets.  Correlations are detected between the signed-intensity of matched features and are examined to determine whether they are due to foreground contamination or instrumental systematics, or whether they are indeed induced by the \isw\ effect.

\subsection{Detections}

\begin{figure}
\centering
\includegraphics[width=\statplotwidth]{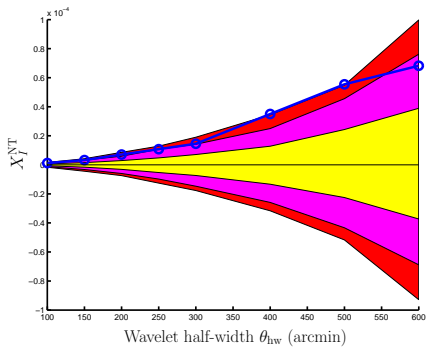}
\caption{Correlation statistics computed for signed-intensity in the matched intensity analysis, from the \wmap\ co-added map and the \nvss\ map. Significance levels obtained from the 1000 Monte Carlo simulations are shown by the shaded regions for 68\% (yellow/light-grey), 95\% (magenta/grey) and 99\% (red/dark-grey) levels.}
\label{fig:stat_intensity}
\end{figure}

The correlation statistics computed from the \wmap\ and \nvss\ data for the signed-intensity of matched features are displayed in \fig{\ref{fig:stat_intensity}}.  Significance levels computed from the Monte Carlo simulations are again shown on each plot.  A non-zero correlation signal is clearly present.  A strong detection at 99.2\% significance is made in the correlation of signed-intensities of local features for wavelet half-width $\halfw\simeq400\arcmin$.  This correlation deviates from the mean of the Monte Carlo simulation by 2.6 standard deviations.
Note that these significance measures are again based on an \emph{a posteriori} scale selection.

We also compute the \emph{a priori} significance of the detection of correlation when considering all scales in aggregate, taking into account the correlation between scales.  This is based on the \chisqd\ test described in \sectn{\ref{sec:morph_detections}}.  For this case we detect a highly significant correlation at a level of 99.9\%.  
%

Although it is possible to localise regions on the sky that contribute most significantly to the correlation detected, we again do not do this since the \isw\ signal is expected to be weakly distributed over the entire sky (as described in \sectn{\ref{sec:morph_detections}}). Instead, we display in \fig{\ref{fig:wcoeff_maps}} the signed-intensity of matched features on the sky for the scale ($\halfw\simeq400\arcmin$) corresponding to the most significance detection of correlation using this analysis.  Due to the strength of the correlation, it is again possible to see it by eye between the \wmap\ (\fig{\ref{fig:wcoeff_maps}}~(b)) and \nvss\ data (\fig{\ref{fig:wcoeff_maps}}~(c)).

\subsection{Foregrounds and systematics}

\begin{figure}
\centering
\subfigure[Individual band maps]{
\includegraphics[width=\statplotwidth]{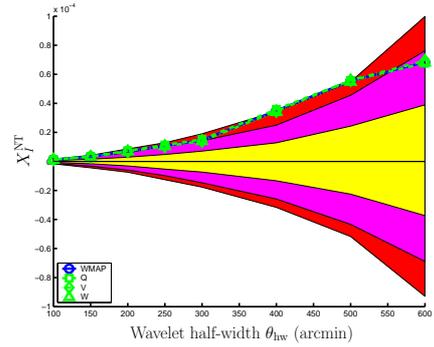}
}\\
\subfigure[Difference maps]{
\includegraphics[width=\statplotwidth]{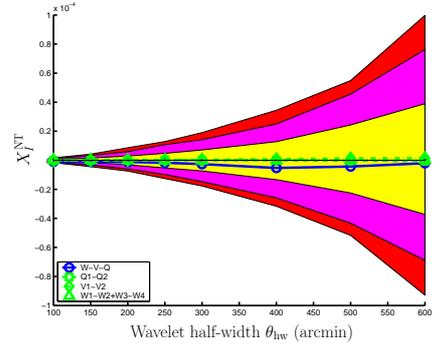}
}
\caption{Correlation statistics computed for the signed-intensity in the matched intensity analysis, from the \wmap\ individual band and difference maps and the \nvss\ map. Significance levels obtained from the 1000 Monte Carlo simulations are shown by the shaded regions for 68\% (yellow/light-grey), 95\% (magenta/grey) and 99\% (red/dark-grey) levels.}
\label{fig:stat_intensity_test}
\end{figure}

For this analysis procedure we also check whether the correlation detected is perhaps due to foreground contamination or systematics.  We follow the same procedure outlined in \sectn{\ref{sec:morph_foregrounds}}.  The correlation signals computed for the matched intensity analysis using individual \wmap\ bands and differences maps in place of the \wmap\ co-added data are plotted in \fig{\ref{fig:stat_intensity_test}}.
No frequency dependence is observed between the signals computed for the individual band maps (see \fig{\ref{fig:stat_intensity_test}~(a)}), neither is any correlation observed in the foreground contaminated W$-$V$-$Q map (see \fig{\ref{fig:stat_intensity_test}~(b)}).  These findings suggest that foreground contributions are not responsible for the correlation detected in this analysis.
Furthermore, the same correlation signal is observed in each individual band (see \fig{\ref{fig:stat_intensity_test}~(a)}), but is not present in any of the difference maps for each given band (see \fig{\ref{fig:stat_intensity_test}~(b)}).  This indicates that systematics are not responsible for the correlation detected in this analysis.
These results again strongly suggest that the \isw\ effect is responsible for the correlation detected between the \wmap\ and \nvss\ data in this second analysis based on the signed-intensity of matched features.

\section{Conclusions}
\label{sec:conclusions}

We have presented the first direct analysis of local morphological measures on the sphere defined through a steerable wavelet analysis.  Using the wavelet on the sphere constructed from the second derivative of a Gaussian, we are able to define three quantities to characterise the morphology of local features: the signed-intensity, orientation and elongation.  These local morphological measures provide additional probes to search for correlation between the \cmb\ and \lss\ of the Universe that may have been induced by the \isw\ effect.

Based on the morphological measures defined, we perform two distinct analyses to search for 
correlation between the \wmap\ and \nvss\ data.  The first procedure, the local morphological analysis, correlates the morphological measures of local features extracted independently in the two data sets.  This provides the most general analysis and does not make any assumption regarding possible correlation in the data.  The second procedure, the matched intensity analysis, is based on the assumption that local features of the \lss\ are somehow included in the local features of the \cmb.
Local features are extracted from the \wmap\ data that are matched in orientation to features in the \nvss\ data and the signed-intensities of these features are correlated between the two data sets.
Although the two analysis procedures performed are obviously different conceptually, we have shown that they differ practically also.  Regardless of the procedure adopted, in the absence of an \isw\ effect the \wmap\ and \nvss\ data should be independent and no correlation should be present in any of the local morphological measures considered.  However, strong detections are obtained in the correlation of the signed-intensities of local features using both analysis procedures.  The most significant detection is made in the matched intensity analysis at 99.9\% significance, when all scales are considered in aggregate.
Moreover, in the local morphological analysis, moderate detections are also made in the correlation of the orientation and elongation of local features.
In both analyses foreground contamination and instrumental systematics have been ruled out as the source of the correlation observed.  This strongly suggests that the correlation detected is indeed due to the \isw\ effect.  Since the \isw\ effect only exists in a universe with dark energy or a non-flat universe, and strong constraints have been placed on the flatness of the Universe \citep{spergel:2006}, the detection of the \isw\ effect made here can be inferred as direct and independent evidence for dark energy.


In previous wavelet analyses to detect the \isw\ effect, theoretical predictions are derived for the wavelet correlation in order to constrain dark energy parameters of cosmological models \citep{vielva:2005,mcewen:2006:isw,pietrobon:2006}.  
As already discussed, our detections give a new insight on the nature of the correlation between the \cmb\ and the \lss, in terms of the signed-intensity, orientation and elongation of local features.  This might in turn help to better understand the nature of dark energy through new constraints on dark energy parameters.
However, our steerable wavelet analysis is complicated by allowing the orientation of wavelet coefficients to vary as a function of the data.  The derivation of the theoretical correlation for each local morphological measure is not easily tractable.  We are currently exploring this issue in more detail and leave any attempts to constrain standard dark energy parameters to a future work.  We also intend to explore the use of this, and other wavelet-based, \isw\ detections to investigate perturbations in the dark energy fluid and to constrain the sound speed of dark energy (see \eg\ \citealt{weller:2003,bean:2004}).

\section*{Acknowledgements}

We thank Patricio Vielva and Enrique Mart\'{\i}nez-Gonz\'{a}lez for providing the preprocessed \nvss\ data.  JDM is supported by PPARC.  YW acknowledges support of the Swiss National Science Foundation (SNF) under contract No. 200021-107478/1. He is also postdoctoral researcher of the Belgian National Science Foundation (FNRS).  Some of the results in this paper have been derived using the \healpix\ package \citep{gorski:2005}.  We acknowledge the use of the \lambdaarchtext\ (\lambdaarch).  Support for \lambdaarch\ is provided by the NASA Office of Space Science.

\bibliographystyle{mymnras_eprint}
\bibliography{bib}

\begin{thebibliography}{52}
\providecommand{\natexlab}[1]{#1}
\providecommand{\url}[1]{\texttt{#1}}
\providecommand{\urlprefix}{URL }
\providecommand{\eprint}[2][]{\url{#2}}

\bibitem[{Afshordi(2004)}]{afshordi:2004}
Afshordi N., 2004, Phys.\ Rev.\ D., D70, 083536,
  \href{http://arXiv.org/abs/astro-ph/0401166}{{\tt astro-ph/0401166}}

\bibitem[{Afshordi et~al.(2004)Afshordi, Loh \& Strauss}]{afshordietal:2004}
Afshordi N., Loh Y.S., Strauss M.A., 2004, Phys.\ Rev.\ D., D69, 083524,
  \href{http://arXiv.org/abs/astro-ph/0308260}{{\tt astro-ph/0308260}}

\bibitem[{{Allen} et~al.(2002){Allen}, {Schmidt} \& {Fabian}}]{allen:2002}
{Allen} S.W., {Schmidt} R.W., {Fabian} A.C., 2002, Mon.\ Not.\ Roy.\ Astron.\
  Soc., 334, L11, \href{http://arXiv.org/abs/astro-ph/0205007}{{\tt
  astro-ph/0205007}}

\bibitem[{Antoine \& Vandergheynst(1998)}]{antoine:1998}
Antoine J.P., Vandergheynst P., 1998, J.\ Math.\ Phys., 39, 8, 3987

\bibitem[{{Barreiro} et~al.(2001){Barreiro}, {Mart{\'{\i}}nez-Gonz{\'a}lez} \&
  {Sanz}}]{barreiro:2001}
{Barreiro} R.B., {Mart{\'{\i}}nez-Gonz{\'a}lez} E., {Sanz} J.L., 2001, Mon.\
  Not.\ Roy.\ Astron.\ Soc., 322, 411,
  \href{http://arXiv.org/abs/astro-ph/0009365}{{\tt astro-ph/0009365}}

\bibitem[{{Barreiro} et~al.(1997){Barreiro}, {Sanz},
  {Mart\'{\i}nez-Gonz\'{a}lez}, {Cayon} \& {Silk}}]{barreiro:1997}
{Barreiro} R.B., {Sanz} J.L., {Mart\'{\i}nez-Gonz\'{a}lez} E., {Cayon} L.,
  {Silk} J., 1997, Astrophys.\ J., 478, 1,
  \href{http://arXiv.org/abs/astro-ph/9612114}{{\tt astro-ph/9612114}}

\bibitem[{{Bean} \& {Dor{\'e}}(2004)}]{bean:2004}
{Bean} R., {Dor{\'e}} O., 2004, Phys.\ Rev.\ D., 69, 083503,
  \href{http://arXiv.org/abs/astro-ph/0307100}{{\tt astro-ph/0307100}}

\bibitem[{Bennett et~al.(2003{\natexlab{a}})}]{bennett:2003b}
Bennett C.L., et~al., 2003{\natexlab{a}}, Astrophys.\ J.\ Supp., 148, 97,
  \href{http://arXiv.org/abs/astro-ph/0302208}{{\tt astro-ph/0302208}}

\bibitem[{Bennett et~al.(2003{\natexlab{b}})}]{bennett:2003a}
Bennett C.L., et~al., 2003{\natexlab{b}}, Astrophys.\ J.\ Supp., 148, 1,
  \href{http://arXiv.org/abs/astro-ph/0302207}{{\tt astro-ph/0302207}}

\bibitem[{Boldt(1987)}]{boldt:1987}
Boldt E., 1987, Phys.\ Rep., 146, 215

\bibitem[{Boughn \& Crittenden(2004)}]{boughn:2004}
Boughn S., Crittenden R., 2004, Nature, 427, 45,
  \href{http://arXiv.org/abs/astro-ph/0305001}{{\tt astro-ph/0305001}}

\bibitem[{Boughn \& Crittenden(2005)}]{boughn:2005}
Boughn S., Crittenden R., 2005, Mon.\ Not.\ Roy.\ Astron.\ Soc., 360, 1013,
  \href{http://arXiv.org/abs/astro-ph/0408242}{{\tt astro-ph/0408242}}

\bibitem[{Boughn \& Crittenden(2002)}]{boughn:2002}
Boughn S.P., Crittenden R.G., 2002, Phys.\ Rev.\ Lett., 88, 021302,
  \href{http://arXiv.org/abs/astro-ph/0111281}{{\tt astro-ph/0111281}}

\bibitem[{{Cabr{\'e}} et~al.(2007){Cabr{\'e}}, {Fosalba}, {Gazta{\~n}aga} \&
  {Manera}}]{cabre:2007}
{Cabr{\'e}} A., {Fosalba} P., {Gazta{\~n}aga} E., {Manera} M., 2007, Mon.\
  Not.\ Roy.\ Astron.\ Soc., 381, 1347,
  \href{http://arXiv.org/abs/astro-ph/0701393}{{\tt astro-ph/0701393}}

\bibitem[{{Cabr{\'e}} et~al.(2006){Cabr{\'e}}, {Gazta{\~n}aga}, {Manera},
  {Fosalba} \& {Castander}}]{cabre:2006}
{Cabr{\'e}} A., {Gazta{\~n}aga} E., {Manera} M., {Fosalba} P., {Castander} F.,
  2006, Mon.\ Not.\ Roy.\ Astron.\ Soc., 372, L23,
  \href{http://arXiv.org/abs/astro-ph/0603690}{{\tt astro-ph/0603690}}

\bibitem[{Condon et~al.(1998)Condon, Cotton, Greisen, Yin, Perley, Taylor \&
  Broderick}]{condon:1998}
Condon J.J., Cotton W.D., Greisen E.W., Yin Q.F., Perley R.A., Taylor G.B.,
  Broderick J.J., 1998, Astrophys.\ J., 115, 1693

\bibitem[{Corasaniti et~al.(2005)Corasaniti, Giannantonio \&
  Melchiorri}]{corasaniti:2005}
Corasaniti P.S., Giannantonio T., Melchiorri A., 2005, Phys.\ Rev.\ D., D71,
  123521, \href{http://arXiv.org/abs/astro-ph/0504115}{{\tt astro-ph/0504115}}

\bibitem[{Crittenden \& Turok(1996)}]{crittenden:1996}
Crittenden R.G., Turok N., 1996, Phys.\ Rev.\ Lett., 76, 575,
  \href{http://arXiv.org/abs/astro-ph/9510072}{{\tt astro-ph/9510072}}

\bibitem[{Fosalba \& Gazta$\tilde{\rm n}$aga(2004)}]{fosalba:2004}
Fosalba P., Gazta$\tilde{\rm n}$aga E., 2004, Mon.\ Not.\ Roy.\ Astron.\ Soc.,
  350, L37, \href{http://arXiv.org/abs/astro-ph/0305468}{{\tt
  astro-ph/0305468}}

\bibitem[{Fosalba et~al.(2003)Fosalba, Gazta$\tilde{\rm n}$aga \&
  Castander}]{fosalba:2003}
Fosalba P., Gazta$\tilde{\rm n}$aga E., Castander F., 2003, Astrophys.\ J.,
  597, L89, \href{http://arXiv.org/abs/astro-ph/0307249}{{\tt
  astro-ph/0307249}}

\bibitem[{{Gazta{\~n}aga} et~al.(2006){Gazta{\~n}aga}, {Manera} \&
  {Multam{\"a}ki}}]{gaztanaga:2006}
{Gazta{\~n}aga} E., {Manera} M., {Multam{\"a}ki} T., 2006, Mon.\ Not.\ Roy.\
  Astron.\ Soc., 365, 171, \href{http://arXiv.org/abs/astro-ph/0407022}{{\tt
  astro-ph/0407022}}

\bibitem[{{Giannantonio} et~al.(2006)}]{giannantonio:2006}
{Giannantonio} T., et~al., 2006, Phys.\ Rev.\ D., 74, 6, 063520,
  \href{http://arXiv.org/abs/astro-ph/0607572}{{\tt astro-ph/0607572}}

\bibitem[{G\'{o}rski et~al.(2005)G\'{o}rski, Hivon, Banday, Wandelt, Hansen,
  Reinecke \& Bartelmann}]{gorski:2005}
G\'{o}rski K.M., Hivon E., Banday A.J., Wandelt B.D., Hansen F.K., Reinecke M.,
  Bartelmann M., 2005, Astrophys.\ J., 622, 759,
  \href{http://arXiv.org/abs/astro-ph/0409513}{{\tt astro-ph/0409513}}

\bibitem[{{Gurzadyan} et~al.(2005)}]{gurzadyan:2005}
{Gurzadyan} V.G., et~al., 2005, Mod.\ Phys.\ Lett.\ A, 20, 813,
  \href{http://arXiv.org/abs/astro-ph/0503103}{{\tt astro-ph/0503103}}

\bibitem[{Hinshaw et~al.(2007)}]{hinshaw:2006}
Hinshaw G., et~al., 2007, Astrophys.\ J.\ Supp., 170, 288,
  \href{http://arXiv.org/abs/astro-ph/0603451}{{\tt astro-ph/0603451}}

\bibitem[{Hu \& Scranton(2004)}]{hu:2004}
Hu W., Scranton R., 2004, Phys.\ Rev.\ D., D70, 123002,
  \href{http://arXiv.org/abs/astro-ph/0408456}{{\tt astro-ph/0408456}}

\bibitem[{Jarrett et~al.(2000)Jarrett, Chester, Cutri, Schneider, Skrutskie \&
  Huchra}]{jarrett:2000}
Jarrett T.H., Chester T., Cutri R., Schneider S., Skrutskie M., Huchra J.P.,
  2000, Astron. J., 119, 2498,
  \href{http://arXiv.org/abs/astro-ph/0004318}{{\tt astro-ph/0004318}}

\bibitem[{Komatsu et~al.(2003)}]{komatsu:2003}
Komatsu E., et~al., 2003, Astrophys.\ J.\ Supp., 148, 119,
  \href{http://arXiv.org/abs/astro-ph/0302223}{{\tt astro-ph/0302223}}

\bibitem[{{LoVerde} et~al.(2007){LoVerde}, {Hui} \& {Gaztanaga}}]{loverde:2006}
{LoVerde} M., {Hui} L., {Gaztanaga} E., 2007, Phys.\ Rev.\ D., 75, 043519,
  \href{http://arXiv.org/abs/astro-ph/0611539}{{\tt astro-ph/0611539}}

\bibitem[{Maddox et~al.(1990)Maddox, Efstathiou, Sutherland \&
  Loveday}]{maddox:1990}
Maddox S.J., Efstathiou G., Sutherland W.J., Loveday J., 1990, Mon.\ Not.\
  Roy.\ Astron.\ Soc., 242, 43

\bibitem[{McEwen et~al.(2007{\natexlab{a}})McEwen, Hobson, Mortlock \&
  Lasenby}]{mcewen:2006:fcswt}
McEwen J.D., Hobson M.P., Mortlock D.J., Lasenby A.N., 2007{\natexlab{a}}, IEEE
  Trans.\ Sig.\ Proc., 55, 2, 520,
  \href{http://arXiv.org/abs/astro-ph/0506308}{{\tt astro-ph/0506308}}

\bibitem[{McEwen et~al.(2007{\natexlab{b}})McEwen, Vielva, Hobson,
  Mart\'{\i}nez-Gonz\'{a}lez \& Lasenby}]{mcewen:2006:isw}
McEwen J.D., Vielva P., Hobson M.P., Mart\'{\i}nez-Gonz\'{a}lez E., Lasenby
  A.N., 2007{\natexlab{b}}, Mon.\ Not.\ Roy.\ Astron.\ Soc., 373, 1211,
  \href{http://arXiv.org/abs/astro-ph/0602398}{{\tt astro-ph/0602398}}

\bibitem[{{Monteser{\'{\i}}n} et~al.(2005){Monteser{\'{\i}}n}, {Barreiro},
  {Sanz} \& {Mart{\'{\i}}nez-Gonz{\'a}lez}}]{monteserin:2005}
{Monteser{\'{\i}}n} C., {Barreiro} R.B., {Sanz} J.L.,
  {Mart{\'{\i}}nez-Gonz{\'a}lez} E., 2005, Mon.\ Not.\ Roy.\ Astron.\ Soc.,
  360, 9, \href{http://arXiv.org/abs/astro-ph/0511308}{{\tt astro-ph/0511308}}

\bibitem[{Nolta et~al.(2004)}]{nolta:2004}
Nolta M.R., et~al., 2004, Astrophys.\ J., 608, 10,
  \href{http://arXiv.org/abs/astro-ph/0305097}{{\tt astro-ph/0305097}}

\bibitem[{Padmanabhan et~al.(2005)Padmanabhan, Hirata, Seljak, Schlegel,
  Brinkmann \& Schneider}]{padmanabhan:2004}
Padmanabhan N., Hirata C.M., Seljak U., Schlegel D.J., Brinkmann J., Schneider
  D.P., 2005, Phys.\ Rev.\ D., D72, 043525,
  \href{http://arXiv.org/abs/astro-ph/0410360}{{\tt astro-ph/0410360}}

\bibitem[{Perlmutter et~al.(1999)}]{perlmutter:1999}
Perlmutter S., et~al., 1999, Astrophys.\ J., 517, 565,
  \href{http://arXiv.org/abs/astro-ph/9812133}{{\tt astro-ph/9812133}}

\bibitem[{Pietrobon et~al.(2006)Pietrobon, Balbi \& Marinucci}]{pietrobon:2006}
Pietrobon D., Balbi A., Marinucci D., 2006, Phys.\ Rev.\ D., 74, 4, 043524,
  \href{http://arXiv.org/abs/astro-ph/0606475}{{\tt astro-ph/0606475}}

\bibitem[{{P}lanck collaboration(2005)}]{planck:bluebook}
{P}lanck collaboration, 2005, {ESA} {P}lanck blue book, Technical Report
  ESA-SCI(2005)1, ESA, \href{http://arXiv.org/abs/astro-ph/0604069}{{\tt
  astro-ph/0604069}}

\bibitem[{Pogosian(2005)}]{pogosian:2004}
Pogosian L., 2005, {JCAP}, 0504, 015,
  \href{http://arXiv.org/abs/astro-ph/0409059}{{\tt astro-ph/0409059}}

\bibitem[{{Pogosian}(2006)}]{pogosian:2006}
{Pogosian} L., 2006, New Astron. Rev., 50, 932,
  \href{http://arXiv.org/abs/astro-ph/0606626}{{\tt astro-ph/0606626}}

\bibitem[{Pogosian et~al.(2005)Pogosian, Corasaniti, Stephan-Otto, Crittenden
  \& Nichol}]{pogosian:2005}
Pogosian L., Corasaniti P.S., Stephan-Otto C., Crittenden R., Nichol R., 2005,
  Phys.\ Rev.\ D., D72, 103519,
  \href{http://arXiv.org/abs/astro-ph/0506396}{{\tt astro-ph/0506396}}

\bibitem[{{Rassat} et~al.(2006){Rassat}, {Land}, {Lahav} \&
  {Abdalla}}]{rassat:2006}
{Rassat} A., {Land} K., {Lahav} O., {Abdalla} F.B., 2006, ArXiv,
  \href{http://arXiv.org/abs/astro-ph/0610911}{{\tt astro-ph/0610911}}

\bibitem[{{Riess} et~al.(1998)}]{riess:1998}
{Riess} A.G., et~al., 1998, Astron.\ J., 116, 1009,
  \href{http://arXiv.org/abs/astro-ph/9805201}{{\tt astro-ph/9805201}}

\bibitem[{Sachs \& Wolfe(1967)}]{sachs:1967}
Sachs R.K., Wolfe A.M., 1967, Astrophys.\ J., 147, 73

\bibitem[{{Scranton} et~al.(2003)}]{scranton:2003}
{Scranton} R., et~al., 2003, ArXiv,
  \href{http://arXiv.org/abs/astro-ph/0307335}{{\tt astro-ph/0307335}}

\bibitem[{Spergel et~al.(2007)}]{spergel:2006}
Spergel D.N., et~al., 2007, Astrophys.\ J.\ Supp., 170, 377,
  \href{http://arXiv.org/abs/astro-ph/0603449}{{\tt astro-ph/0603449}}

\bibitem[{{Vielva} et~al.(2006){Vielva}, {Mart{\'{\i}}nez-Gonz{\'a}lez} \&
  {Tucci}}]{vielva:2005}
{Vielva} P., {Mart{\'{\i}}nez-Gonz{\'a}lez} E., {Tucci} M., 2006, Mon.\ Not.\
  Roy.\ Astron.\ Soc., 365, 891,
  \href{http://arXiv.org/abs/astro-ph/0408252}{{\tt astro-ph/0408252}}

\bibitem[{Weller \& Lewis(2003)}]{weller:2003}
Weller J., Lewis A.M., 2003, Mon.\ Not.\ Roy.\ Astron.\ Soc., 346, 987,
  \href{http://arXiv.org/abs/astro-ph/0307104}{{\tt astro-ph/0307104}}

\bibitem[{Wiaux et~al.(2005)Wiaux, Jacques \& Vandergheynst}]{wiaux:2005}
Wiaux Y., Jacques L., Vandergheynst P., 2005, Astrophys.\ J., 632, 15,
  \href{http://arXiv.org/abs/astro-ph/0502486}{{\tt astro-ph/0502486}}

\bibitem[{Wiaux et~al.(2006)Wiaux, Jacques, Vielva \&
  Vandergheynst}]{wiaux:2005c}
Wiaux Y., Jacques L., Vielva P., Vandergheynst P., 2006, Astrophys.\ J., 652,
  820, \href{http://arXiv.org/abs/astro-ph/0508516}{{\tt astro-ph/0508516}}

\bibitem[{Wiaux et~al.(invited contribution, 2007)Wiaux, McEwen \&
  Vielva}]{wiaux:2006:review}
Wiaux Y., McEwen J.D., Vielva P., invited contribution, 2007, J.\ Fourier
  Anal.\ and Appl., 13, 4, 477,
  \href{http://arXiv.org/abs/arXiv:0704.3144}{{\tt arXiv:0704.3144}}

\bibitem[{{York D.~G. {\it et al.} (SDSS collaboration)}(2000)}]{york:2000}
{York D.~G. {\it et al.} (SDSS collaboration)}, 2000, Astron.\ J., 120, 1579,
  \href{http://arXiv.org/abs/astro-ph/0006396}{{\tt astro-ph/0006396}}

\end{thebibliography}

\label{lastpage}
\end{document}